\documentclass[cits]{PoS}

\usepackage[clockwise]{rotating}
\usepackage{epsfig}

\newenvironment{myitemize}{ 
	\begin{list}{$\bullet$}{
\setlength{\itemsep}{0pt}
\setlength{\topsep}{0pt}
}
}{
	\end{list}
}

\newenvironment{myitemizei}{ 
	\begin{list}{$\bullet$}{
\setlength{\itemsep}{-4pt}
\setlength{\topsep}{-6pt}
}
}{
	\end{list}
}

\newenvironment{myitemizeii}{ 
	\begin{list}{$\bullet$}{
\setlength{\itemsep}{0pt}
\setlength{\topsep}{6pt}
}
}{
	\end{list}
}

\def\be{\begin{equation}}
\def\ee{\end{equation}}
\def\bea{\begin{eqnarray}}
\def\eea{\end{eqnarray}}

\def\good{$\star\star\star$}
\def\soso{$\star\star$}
\def\bad{$\star$}

\def\reff#1{\ref{#1}}

\def\eq#1{Eq.~(\reff{#1})}

\def\fig#1{Fig.~\reff{#1}}

\def\sec#1{Sec.~\reff{#1}}
\def\tab#1{Table~\reff{#1}}

\newcommand{\lsim}{ {\
\lower-1.2pt\vbox{\hbox{\rlap{$<$}\lower5pt\vbox{\hbox{$\sim$}}}}\ } }
\newcommand{\gsim}{ {\
\lower-1.2pt\vbox{\hbox{\rlap{$>$}\lower5pt\vbox{\hbox{$\sim$}}}}\ } }

\def\er#1#2{\relax\ifmmode{}^{+#1}_{-#2}\else$^{+#1}_{-#2}$\fi}
\def\erparen#1#2{\relax\ifmmode{}(^{#1}_{#2})\else$(^{#1}_{#2})$\fi}
\def~{\ifmmode\phantom{0}\else\penalty10000\ \fi}

\def\fm{\mathrm{fm}}
\def\ev{\mathrm{e\kern-0.1em V}}
\def\kev{\mathrm{ke\kern-0.1em V}}
\def\mev{\mathrm{Me\kern-0.1em V}}
\def\gev{\mathrm{Ge\kern-0.1em V}}
\def\tev{\mathrm{Te\kern-0.1em V}}

\def\re{\,\mathrm{Re}}
\def\im{\,\mathrm{Im}}

\def\n#1e#2n{{#1}\times 10^{#2}}

\def\ra{\rangle}
\def\la{\langle}

\def\bea{\begin{eqnarray}}
\def\eea{\end{eqnarray}}
\def\nn{\nonumber}

\def\cH{\mathcal{H}}

\def\ods2{\mathcal{O}_{\Delta S=2}}
\def\zds2{Z_{\Delta S=2}}

\def\msbar{{\overline{\mathrm{MS}}}}

\makeatletter
\def\slash#1{{\mathpalette\c@ncel{#1}}} 
\def\big#1{{\hbox{$\left#1\vbox to1.012\ht\strutbox{}\right.\n@space$}}}
\def\Big#1{{\hbox{$\left#1\vbox to1.369\ht\strutbox{}\right.\n@space$}}}
\def\bigg#1{{\hbox{$\left#1\vbox to1.726\ht\strutbox{}\right.\n@space$}}}
\def\Bigg#1{{\hbox{$\left#1\vbox
to2.083\ht\strutbox{}\right.\n@space$}}}
\makeatother

\title{Kaon physics: a lattice perspective}

\ShortTitle{Kaon physics}

\author{\speaker{Laurent Lellouch}\\
        Centre de Physique Th\'eorique, Case 907, CNRS Luminy, F-13288
        Marseille Cedex 9, France~\thanks{CPT is UMR 6207 of 
the CNRS and of
    the universities of Aix-Marseille I, Aix-Marseille II and of Sud Toulon-Var,
and is affiliated with the FRUMAM.}\\
        E-mail: \email{lellouch@cpt.univ-mrs.fr}}

\abstract{I critically review recent lattice QCD results relevant for kaon
  phenomenology, as well as the methods that are used to obtain them. The
  focus is on calculations with $N_f=2$ and $N_f=2+1$ flavors of sea
  quarks. Concerning methodology, the subjects covered include a discussion of
  how best to extrapolate and/or interpolate results to the physical
  quark-mass point, a scheme for assessing the extent to which a lattice QCD
  calculation includes the various effects required to compute a given
  quantity reliably and a procedure for averaging lattice results. The
  phenomenological topics that I review comprise leptonic and semileptonic
  kaon decays, as well as neutral kaon mixing and CP violation in $K\to\pi\pi$
  decays.}

\FullConference{The XXVI International Symposium on Lattice Field Theory\\
		 July 14-19 2008\\
		 Williamsburg, Virginia, USA}

\begin{document}

\section{Introduction}

This talk critically reviews recent lattice QCD results relevant for kaon
phenomenology, as well as the methodology that is used to obtain them. The
focus is on full QCD calculations, which account for the effects of light sea
quarks either partially, as in $N_f=2$ simulations, where degenerate up and
down sea quarks of mass $m_{ud}$ are included, or fully, as in $N_f=2+1$
calculations, where strange sea quarks of mass $m_s$ are also incorporated.

The main motivation for studying kaon physics off and on the lattice is to
test the standard model, to determine some of its parameters and to constrain
possible new physics scenarios. From a lattice perspective, kaon processes
fall into three broad categories. The first are processes, such as leptonic
and semileptonic kaon decays, for which lattice QCD methods are already
providing high precision results.  The second category corresponds to
processes for which lattice calculations are delivering results with errors
on the level of 10\% or less, such as for $K^0$-$\bar K^0$ mixing matrix
elements. The last category of processes are those for which lattice
calculations have failed up until now to provide reliable answers. Amongst
them are the $\Delta I=1/2$ rule and, more critically, direct CP violation in
$K\to\pi\pi$ decays.

Another motivation for studying kaons physics on the lattice is the overlap
this physics has with chiral perturbation theory (ChPT). ChPT describes the
low-energy dynamics of the pseudo-Nambu-Goldstone bosons of chiral symmetry
breaking and has been successful in many phenomenological applications.
Moreover, it is a very useful tool for understanding the dependence of lattice
results on light quark masses and on volume. Recent $N_f=2$ and $2+1$
calculations, which include pions with masses $M_\pi\lsim 350\,\mev$, are not
only using ChPT but are also beginning to provide information about ChPT in
return.

The talk begins with a critical discussion of the role that ChPT and other
expansions can play in interpolating and extrapolating lattice QCD results to
the physical mass point, $(m_{ud},m_s)=(m_{ud}^{ph},m_s^{ph})$, in view of the
quark mass values currently reached in lattice calculations. In an aside, I
present a scheme for visualizing the extent to which a lattice calculation
includes the different effects necessary for computing a quantity of
interest reliably, and a procedure for averaging lattice results. This is
followed by a review of calculations of quantities relevant for leptonic and
semileptonic kaon decays, as well as for neutral kaon mixing and CP violation
in $K\to\pi\pi$ decays. 

\section{Reaching the physical mass point}

Using today's algorithms, it is straightforward to perform $N_f=2+1$
calculations with a strange quark whose mass is around its physical value. The
physical strange quark mass point is thus recovered simply by interpolation.

Reaching the physical up and down quark mass point is much more
difficult. Though the results of PACS-CS~\cite{Aoki:2008sm} announce that
calculations will soon be done directly at this point in physically large
volumes, for the moment all other simulations are being performed with larger
quark masses. Thus, reaching the physical point still requires conducting a
number of computationally intensive calculations with $m_{ud}<m_{ud}^{max}\sim
m_s^{ph}/2$, extending preferably below $m_s^{ph}/12$, and performing a delicate
extrapolation in $m_{ud}$ to $m_{ud}^{ph}\simeq m_s^{ph}/26$.

To guide the interpolation to $m_s^{ph}$ and extrapolation to
$m_{ud}^{ph}$, a natural candidate is $SU(3)$ ChPT, since it provides a concise
framework for describing the dependence of hadronic quantities on the masses
of the up, down and strange quarks. Moreover, ChPT in its various quenched and
partially-quenched guises has served the lattice community well. Nevertheless,
lattice calculations are reaching regions of parameter space and precisions
never attained before, and it is worth considering the following two questions
candidly:
\begin{itemize}

\item What is the best way to interpolate to $m_s=m_s^{ph}$?

\item What is the best way to extrapolate from $m_s^{ph}/12\lsim
m_{ud}< m_{ud}^{max}\sim
m_s^{ph}/2$ to $m_{ud}=m_{ud}^{ph}$?

\end{itemize}
There are, I believe, three physically motivated options to choose from:
\begin{itemize}

\item[(1)] As already mentioned, $SU(3)$ ChPT is a natural candidate. It has
  the advantage of addressing both problems together, within a compact and
  constrained framework. Its drawback is that it provides similar solutions
  to two problems which are of a quite different nature: the first concerns a
  simple interpolation rather far away from the chiral point while the second
  involves a difficult extrapolation which reaches much deeper into the chiral
  regime.

\item[(2)] $SU(2)$ ChPT provides a means of distinguishing these two
  problems. For the extrapolation in $m_{ud}$, it brings to bear all of the
  power of chiral expansions. The interpolation in $m_s$ is not directly
  addressed, but it suffices to supplement the chiral expansion with a regular
  mass--or what I call {\em ``flavor''}--expansion about $m_s^{ph}$, and to
  perform a simple polynomial interpolation.

\item[(3)] The idea of a {\em flavor} expansion can also be applied to the
  extrapolation in $m_{ud}$. To reduce uncertainties, this expansion should be
  performed about the midpoint of the interval between the physical point and
  the largest up and down quark mass considered, i.e. $\bar
  m_{ud}=[m_{ud}^{ph}+ m_{ud}^{max}]/2$. In this scheme, both the
  extrapolation in $m_{ud}$ and the interpolation in $m_s$ can be performed
  with polynomial {\em flavor} expansions.

\end{itemize}
Let us now review these three alternatives in more detail.

\subsection{$SU(3)$ versus $SU(2)$ ChPT and {\em flavor} expansions: 
what's the difference?}

The {\em flavor} expansions are performed about regular points $\bar m_{ud}$
and $m_s^{ph}$ (i.e.\ they are Taylor expansions). This is not the case for the
chiral expansions. $SU(2)$ ChPT is an expansion about the singular point
$(m_{ud},m_s)=(0,m_s^{ph})$.  $SU(3)$ ChPT makes the additional assumption
that the strange quark is chiral so that the expansion is around
$(m_{ud},m_s)=(0,0)$.

In {\em flavor} expansions of quantities which do not vanish in the $SU(2)$
chiral limit, it is the ``distance'' from the expansion points, $\bar m_{ud}$
or $m_s^{ph}$, in units of the QCD scale, which determines how well the series
converges (hence my use of the adjective ``flavor''). Thus, the expansion
parameters are $(m_{ud}-\bar m_{ud})/M_{QCD}$ and $(m_s-m_s^{ph})/M_{QCD}$,
where $M_{QCD}\sim 1\,\gev$ is a typical QCD scale. On the other hand, $SU(3)$
ChPT expressions are expansions in $m_{ud,s}/\Lambda_\chi$, with
$\Lambda_\chi\sim 4\pi F_\pi=O(M_{QCD})$ the chiral symmetry breaking
scale. In $SU(2)$, the expansions are in $m_{ud}/m_s$ and
$m_{ud}/\Lambda_\chi$.

Because $m_{ud}$ and $m_s$ are not measured directly in experiment, it is
convenient to replace these masses by observables which are sensitive to
them. ChPT suggests that $M_\pi$ and
$M_K^{\chi}\equiv[M_K^2-M_\pi^2/2]^{1/2}$, with $M_\pi^{ph}\simeq 135\,\mev$
and $M_K^{\chi,ph}\simeq 486\,\mev$, are particularly appropriate. Indeed, LO
ChPT yields $M_\pi^2=2Bm_{ud}$ and $(M_K^{\chi})^2=Bm_s $, with $B=
O(M_{QCD})$. In terms of these variable, the $SU(2)$ ChPT expansion parameters
can be written $(M_{\pi}/\sqrt2 M_K^\chi)^2$ and $(M_{\pi}/\Lambda_\chi)^2$,
while $SU(3)$ ChPT is an expansion in
$(M_{\pi,K,\eta}/\Lambda_\chi)^2$. Similarly, the {\em flavor} expansion
parameters become $\Delta_\pi\equiv (M_\pi^2-\bar M_\pi^2)/2M_{QCD}^2$ and
$\Delta_K\equiv [(M_K^{\chi})^2-(M_K^{\chi,ph})^2]/M_{QCD}^2$. It is worth
noting that this definition for $\Delta_K$ remains appropriate if one assumes
that $M_K^\chi$ itself obeys a {\em flavor} expansion in $m_s$,
i.e. $M_K^{\chi}=M_K^{\chi,ph}[1+C_K (m_s-m_s^{ph})/M_{QCD}+\mbox{h.o.t}]$,
with $C_K$ a constant. Indeed, in that case we also have
$\Delta_K=O[(m_s-m_s^{ph})/M_{QCD}]$. On the other hand, $M_\pi$'s {\em
  flavor} expansion in $m_{ud}$, $M_\pi=\bar M_\pi[1+2C_\pi (m_{ud}-\bar
  m_{ud})/M_{QCD}+\mbox{h.o.t}]$, is poorly behaved for the range of $m_{ud}$
currently considered in lattice calculations, since the NLO plus higher order
terms can be 50\% or more of the LO term. However, this fact does not
invalidate the use of {\em flavor} expansions in $m_{ud}$ for quantities which
do not vanish in the $SU(2)$ chiral limit. It merely signals that, in current
calculations, the relative variation in $M_\pi$ is large, while the change in
$M_\pi$ with respect to $M_{QCD}$ remains small.

The expected accuracy at NLO in the $SU(2)$ expansion around the physical mass
point is much better than for the $SU(3)$ case. Indeed, in $SU(2)$ this
accuracy is given by $(M_{\pi}^{ph}/\sqrt2 M_K^{\chi,ph})^4\sim 0.1\%$ whereas
it is expected to be $(M_\eta^{ph}/4\pi F_\pi)^4\sim 5\%$ in the $SU(3)$
case. However, with pions of about 450~MeV floating around, as in present day
simulations, the $SU(2)$ figure becomes $(M_{\pi}/\sqrt2 M_K^\chi)^4\sim
20\%$, which is much less impressive. Nevertheless, this expansion has the
advantage that its convergence improves rapidly as $M_\pi$ is reduced, while
the $SU(3)$ expansion parameter $(M_\eta/4\pi F_\pi)^2$ does not decrease
significantly with $M_\pi$.

The accuracy of the {\em flavor} interpolation in strange quark mass is
generically very high. Suppose that one has performed the calculation for at
least two values of the strange quark mass that bracket $m_s^{ph}$ with a
total spread of about 10\%. The expansion parameter is then $|\Delta_K|\sim
0.01$. Assuming that the error due to the truncation of the interpolating
polynomial is on the order of the first omitted term, the systematic error
associated with a linear interpolation in $(M_K^{\chi})^2$ (i.e. a linear
interpolation) will have an accuracy on the order of $\Delta_K^2\sim 0.01\%$.

In current lattice calculations, the {\em flavor} expansion in up and down
quark mass is not as good. Assuming that we consider only pions with $M_\pi\le
M_\pi^{max}=450\,\mev$, the expansion parameter is $|\Delta_\pi|\lsim
0.05$. This means that a linear extrapolation will have a truncation
uncertainty on the order of $\Delta_\pi^2{\sim}0.3\%$ (with a coefficient that
increases with $u{/}d$ content). Moreover, it is straightforward to show that,
with a quadratic {\em flavor} expansion, one can fit a chiral logarithm which
gives a correction of up to 30\% as $M_\pi$ varies in the range from
$M_\pi^{ph}$ to $M_\pi^{max}$, with a systematic accuracy better than
0.5\%. So, even in the presence of a chiral logarithm, a {\em flavor}
expansion can be used.

Let me now add a few words about the possible outcomes of implementing the
different approaches.  $SU(3)$ ChPT provides functional forms which are more
constrained, i.e. which have less parameters, at a given order, than the
$SU(2)$ chiral and {\em flavor} expansions. That is one reason why $SU(3)$
ChPT might be appealing. So let me assume, for the moment that we are fitting
lattice results to $SU(3)$ ChPT expressions. As $M_\pi$ is lowered below
$\sqrt2 M_K^\chi$ with fixed $m_s$, $SU(3)$ ChPT turns into $SU(2)$ ChPT,
except that the extended symmetry of the $SU(3)$ theory imposes constraints
amongst the $SU(2)$ LECs. These constraints can be released by adding NNLO and
higher terms to the $SU(3)$ expansion. If the $M_K^2/\Lambda_\chi^2$ expansion
in the $SU(3)$ theory behaves well, then the LECs obtained with the fits may
be $SU(3)$ LECs of QCD, as defined in the $SU(3)$ chiral limit. However, if
the assumption that the strange quark is chiral is not borne out in practice,
a good fit may still be obtained by adding higher order terms, but the fitted
LECs will most likely not be QCD's LECs. In that case, one may still find that
the $M_\pi^2$ component of the $SU(3)$ chiral expansion is reasonably well
behaved. If this is so, an $SU(2)$ chiral fit ought to work and should give
the $SU(2)$ LECs of QCD. However, the expansion may still behave poorly for
heavier pions because in that case the expansion parameter $(M_{\pi}/\sqrt2
M_K^\chi)^2$ may not be small. Alternatively one may use the {\em flavor}
expansion approach.  It deals with the strange quark mass interpolation in the
same way as $SU(2)$ ChPT, but differs in the choice of expansion point for the
extrapolation in $m_{ud}$. ChPT expands observables about $M_\pi=0$, which is
further from the lightest simulated $M_\pi$ than is the physical point. The
{\em flavor} expansion, on the other hand, is performed about a value of
$M_\pi=\bar M_\pi$ which is between the heaviest simulated $M_\pi$ and the
physical value. Thus, the {\em flavor} expansion will be better behaved,
though generically less constrained.

ChPT is a worthy object of study in its own right, with applications which go
beyond present lattice QCD capabilities. Thus, it is important to test its
range of validity and its accuracy where it is applicable. It is also
important to determine its LECs, since these can be used to make predictions
in a variety of processes. However, if the goal is to determine the value of
an observable at the physical point, one should remain agnostic in regards to
the expansion used and pick the one which gives the lowest combined
statistical and systematic error. Moreover, if the goal is to obtain the LECs
of QCD, it may be necessary to perform calculations closer to the chiral
limit, especially in the case of $SU(3)$ ChPT.

\subsection{$SU(3)$ versus $SU(2)$ ChPT and {\em flavor} expansions: examples}

To further clarify the difference between the different expansions and their
applicability to lattice calculations, it is useful to turn to a concrete
example. We consider here the expansions of the
pion and kaon decay constants, $F_\pi$ and $F_K$, at NLO. In the $SU(3)$
theory, we have~\cite{Gasser:1984gg}:
\bea
F_\pi &=& {F_3}\Bigg\{1-\frac1{(4\pi
  {F_3})^2}\Bigg[\chi_1(M_\pi^2)+ \frac12 \chi_1(M_K^2)\Bigg]
+4{\big(L_5+L_4\big)(\mu)}\frac{M_\pi^2}{F_3^2}+ 8{L_4(\mu)}
\frac{M_K^2}{F_3^2}
\Bigg\}\\
F_K &=& {F_3}\Bigg\{1-\frac1{(4\pi
  {F_3})^2}\Bigg[\frac38 \chi_1(M_\pi^2)+ \frac34 \chi_1(M_K^2)
+\frac38 \chi_1(M_\eta^2)\Bigg]
+4{\big(L_5+2L_4\big)(\mu)}\frac{M_K^2}{F_3^2}\nn\\
& & +4{L_4(\mu)}
\frac{M_\pi^2}{F_3^2}\Bigg\}\ ,
\eea
where $\chi_n(M^2)=M^{2n}\ln (M^2/\mu^2)$ and where $F_3$ is the pion decay
constant in the $N_f=3$ chiral limit. The up-down and strange quark
mass-dependence of these two quantities are obtained here in terms of only {\em
  three} parameters: $F_3$, $L_4$ and $L_5$.

The $SU(2)$ theory is much less frugal with parameters. At NLO
it predicts~\cite{Gasser:1983yg,Sharpe:1995qp,Allton:2008pn}:
\bea
F_\pi &=& {F_2}(1+{\alpha_F}\Delta_K)\left\{
1-\frac{1}{(4\pi {F_2})^2}\left[\chi_1(M_\pi^2)-
{\ell_4(\mu)}M_\pi^2\right]\right\} + O\left(M_\pi^2\Delta_K\right)
\\
F_K &=& {F_2^K}(1+{\alpha_F^K}\Delta_K)\left\{
1-\frac{1}{(4\pi {F_2})^2}\left[\frac38\chi_1(M_\pi^2)-
{\ell_4^K(\mu)}M_\pi^2\right]\right\}+ O\left(M_\pi^2\Delta_K\right)
\ ,\eea
where $F_2$ and $F_2^K$ are the pion and kaon decay constants, respectively,
in the $N_f=2$ chiral limit and where I have included a strange quark mass
dependence. Thus, the $SU(2)$ description of the mass-dependence of the two
decay constants requires at least {\em six} parameters
$(F_2,\,\ell_4,\,\alpha_F,\,F_2^K,\, \ell_4^K, \,\alpha_F^K)$, {\em eight} if
$O(M_\pi^2\Delta_K)$ terms are required.

This number of parameters is comparable to that required in the {\em
  flavor} expansion of $F_K$ and $F_\pi$. {\em Six} parameters are needed if
the $M_\pi^2$ dependence turns out to be linear and {\em eight} if curvature
is observed, corresponding to an expansion to $O(\Delta_\pi^2,\Delta_K)$.

\medskip

\begin{figure}[t]
\begin{center}
\psfig{file=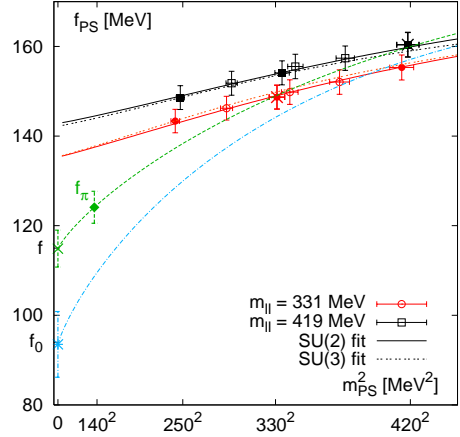,angle=270,width=7.cm}
\caption{\label{fig:FpiRBC08}\em RBC/UKQCD's partially quenched pion decay
  constants versus pion mass squared, for two values of the sea pion mass,
  $M_\pi=331$ and $419\,\mev$. The only points where sea and valence quarks are
  degenerate are the square and circle with crosses. Also shown are the
  unitary $SU(3)$ and $SU(2)$ fits. Conventions are such that $f_\pi=\sqrt2
  F_\pi=131\,\mev$.\vspace{-0.5cm}}
\end{center}
\end{figure}

Let us now investigate how these considerations play out with real lattice
results. I begin with a partially quenched, $N_f=2+1$ study of $F_\pi$ and
$F_K$ performed by RBC/UKQCD~\cite{Allton:2008pn}, whose results were
presented at this conference by E. Scholz~\cite{Scholz:2008uv}. These results
are shown in \fig{fig:FpiRBC08}, where the pion decay constant is plotted
against the valence pion mass squared for two values of the sea pion mass
(331~MeV and 419~MeV). Details of the simulation are given below in
\tab{tab:FKFpi}.

In their calculation, the $SU(3)$ ChPT expansion parameters are, at
$M_\pi^{max}=419\,\mev$: $(M_\pi^{max}/$ $4\pi F_\pi^{ph})^2\simeq 0.1$ and
$(M_\eta/$ $4\pi F_\pi^{ph})^2{\simeq} 0.3$. The $SU(2)$ expansion at
$M_\pi^{max}$ is not any better: $(M_\pi^{max}/$ $\sqrt2
M_K^{\chi,ph})^2\simeq 0.4$. Thus it is not clear, a priori, which of the two
expansions is better at the top of the $M_\pi$ range. Of course, as already
mentioned, as $M_\pi$ decreases the $SU(2)$ expansion improves rapidly whereas
the $SU(3)$ expansion parameter, $(M_{K,\eta}/4\pi F_\pi^{ph})^2$ stays
roughly constant. Assuming that $SU(3)$ ChPT is applicable, they find very
large NLO corrections to the pion decay constant, even at their lightest
unitary point, $M_\pi=311\,\mev$, where they are of order 70\%. They also find
that the NLO forms do not describe their kaon results, where the down quark is
replaced by a strange. This is perhaps not too surprising since their kaons
have masses of up to approximately 570~MeV.

With $SU(2)$ ChPT, on the other hand, they obtain good fits and find much more
reasonable NLO corrections, that are on the order of 30\% at
$M_\pi=311\,\mev$. They use this information, together with that obtained
from fits with partial NNLO terms and more massive pions, to conclude that
$SU(3)$ ChPT fails in the range of masses explored, while $SU(2)$ ChPT is
reliable.

A few comments are in order. The first is that the fits do not take into
account correlations which are obviously strong at fixed sea quark mass. This
makes getting a meaningful figure of merit for the fits difficult. The second
is that the results display none of the logarithmic behavior which becomes
relevant in the extrapolation to physical $M_\pi$: at NLO in partially
quenched ChPT, the dependence on valence quark mass is linear and with only
two values of the sea quark mass, one cannot distinguish between a straight
line and a chiral logarithm. Thus, the lattice results are not inconsistent
with $SU(2)$ ChPT, but they cannot be claimed, either, to confirm the
relevance of this expansion in the quark mass range considered. Moreover, the
value of $F_\pi$ obtained by linear fit would be significantly larger than the
one found in the plot, though consistent within the final systematic error
quoted by the authors. Finally, it should be remembered that the analysis is
performed at a single, rather large value of the lattice spacing ($a\simeq
0.11\,\fm$), and mass dependent discretization errors could distort the
physical chiral behavior.

\begin{figure}[t]
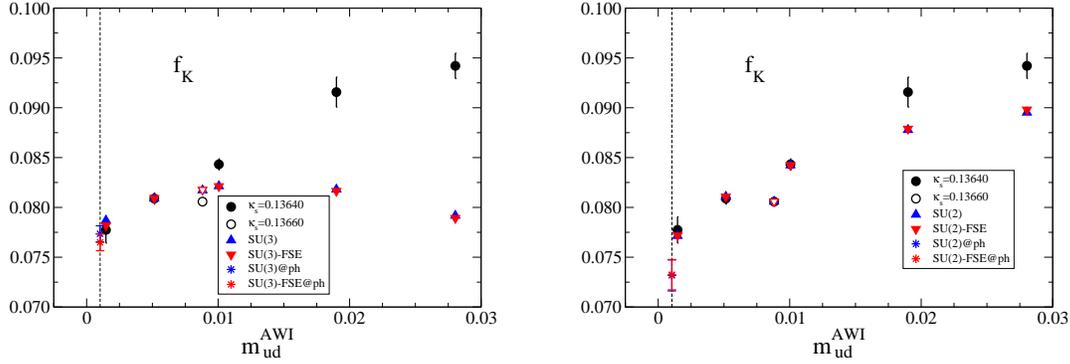

\begin{center}
\psfig{file=fk_SU3,width=6.5cm}
\hspace{1cm}\psfig{file=fk_SU2,width=6.5cm}
\caption{\label{fig:FKPACSCS08} \em From PACS-CS, $F_K$ as a function of the
  isospin averaged up and down quark masses, both in lattice units, at a
  single value of the lattice spacing, $a\simeq
  0.09\,\fm$~\cite{Aoki:2008sm}. The black circles are the decay constants
  obtained from the different simulations, corresponding to $M_\pi\simeq 156$,
  $296$, $385$, $411$, $570$ and $702\,\mev$. The left panel shows these
  decays constants together with results from a variety of NLO $SU(3)$ ChPT
  fits (triangles) while the right panel displays the same results with NLO
  $SU(2)$ ChPT fits (triangles). If all fits were good, triangles at each
  $m_{ud}^{AWI}$ would sit atop the corresponding circle. In their
  conventions, $f_K=\sqrt2 F_K=159\,\mev$.\vspace{-0.7cm}}
\end{center}
\end{figure}

PACS-CS has also investigated the applicability of the two variants of ChPT to
their results for the decay constants~\cite{Aoki:2008sm}, as shown at this
conference by Y. Kuramashi~\cite{Kuramashi:2008tb}. Their calculations are
performed for six different values of the pion mass, ranging from $\sim
700\,\mev$ all the way down to 156~MeV. Moreover, they consider only unitary
points, for which valence and sea quarks of the same flavor have identical
masses (i.e. no partial quenching). The parameters of their simulations are
given below in \tab{tab:FKFpi}. Their studies of the dependence of $F_K$ on
the isospin averaged up and down quark mass, $m_{ud}$, are shown in
\fig{fig:FKPACSCS08}. The left hand panel displays the decay constants
obtained directly from the simulations together with the values of these
constants which result from fitting the simulation data to various $SU(3)$
ChPT forms. The fits are restricted to points with $M_\pi\lsim
410\,\mev$. The fit results above this point are extrapolations. They
find that NLO $SU(3)$ ChPT fails to reproduce the $M_\pi^2$ dependence of
$F_K$ above $M_\pi\sim 400\,\mev$. Moreover, they find that it fails to
predict the strange quark mass dependence of $F_K$ around $m_s^{ph}$ and for
$M_\pi\simeq 400\,\mev$.

Again, the situation is quite different for $SU(2)$ ChPT fits. There they find
that the $m_{ud}$ dependence is well reproduced up to $M_\pi\simeq 410\,\mev$
and only deviates from the simulation result by 5\% at $M_\pi\simeq
570\,\mev$. Moreover the $m_s$ dependence is correctly reproduced, as it
should since there are two $m_s$ values and this dependence is fitted by a line.

These calculations, performed almost all the way down to the physical point,
are a real prowess. For the moment, however, the volume considered for their
lightest pion ($M_\pi\simeq 156\,\mev$) is small, corresponding to $LM_\pi\sim
2.3$. This may make it difficult to control finite-volume effects at low
$M_\pi$. Moreover, the calculations have only been performed at a single
lattice spacing for now, so that alterations of the mass dependence by
discretization errors have not yet been investigated.


Combining the experiences of RBC/UKQCD and PACS-CS, the following conclusion
seems to emerge: $SU(3)$ ChPT appears to break down at the physical strange
quark mass, at least in the presence of heavier up and down quarks, with
masses larger than $m_{ud}\simeq 9 m_{ud}^{ph}\simeq m_s^{ph}/3$,
corresponding to $M_\pi\gsim 400\,\mev$.

\section{Aside on a classification scheme for lattice simulations and on
  the averaging of lattice results}

Before turning to quantities of phenomenological interest, I wish to say a few
words about the methodology that I will follow in reviewing lattice results.

\subsection{Of stars and lattice calculations}
\label{sec:stars}

The FLAVIAnet Lattice Averaging Group (FLAG) is in the process of putting
together a classification scheme for lattice calculations. The goal is to
provide tables which, at a glance, give the reader a sense of how thoroughly a
given calculation includes all of the necessary ingredients, based on a list of
pre-defined, objective criteria. Since this collective work has not yet been
finalized, I propose a personal version of the scheme here.
It is based on a starring system, reminiscent of the one used in a
famous, red restaurant guide:
\begin{itemize}
\item[\good] indicates that this aspect of the calculation is
  fully satisfactory;
\item[\soso] indicates that the corresponding ingredient has not been fully
  included, but that the invstigations performed allow for a reasonable
  estimate of the ensuing systematic error;
\item[\bad] indicates that the calculations performed are not sufficient to
  provide a reliable estimate of what is missing.
\end{itemize}
More specifically, here are the criteria which I use for starring the
calculations reviewed below:
\begin{myitemize}

\item {\bf publication status}
\begin{myitemizei}

\item[\good] published
\item[\soso] preprint
\item[\bad] proceedings, talk

\end{myitemizei}

\item{\bf action, unitarity}

\begin{myitemizei}

\item[\good] local action, unitary calculation
\item[\soso] non-local action and/or discretization induced unitarity
violations
\end{myitemizei}

\item {\bf flavors}

\begin{myitemizei}

\item[\good] all dynamical flavors required for the process under study are
  included
\item[\soso] some dynamical flavors missing, but at least $N_f\ge 2$
\item[\bad] $N_f=0$ (i.e. quenched calculation)

\end{myitemizei}

\item {\bf renormalization}

\begin{myitemizei}

\item[\good] nonperturbative with nonperturbative running 
\item[\soso] nonperturbative with perturbative running at GeV energies, or
  perturbative at two-loops or more
\item[\bad] one-loop perturbative and/or discretization which leads to poorly
  controlled operator mixing

\end{myitemizei}

\item {\bf extrapolation/interpolation to physical mass point}

Let $M_{\pi,typ}^{min}$ be a mass that is representative (``typical'') of the
masses of the lightest pion variants that contribute to the $m_{ud}$
dependence of the quantities studied.~\footnote{\label{foot:mpityp} This
  ``typical'' mass depends on the fermion formulation used, on the quantities
  studied, etc. Since it is meant to be indicative, I have kept its
  determination simple. For staggered calculations I take the RMS of the
  masses of the different tastes; for non-staggered on staggered, the RMS of
  the valence and the sea taste-singlet pion masses; for Wilson, overlap,
  domain-wall, the RMS of the valence and sea pion masses (i.e. simply the
  lightest pion mass for unitary calculations); and for tmQCD, I
  have taken the charged pion mass, though some sort of isospin averaging
  should probably be performed. I thank C. Aubin, J. Laiho, S. Sharpe and
  R. Van de Water for enlightening correspondence.}

\begin{myitemizei}

\item[\good] $M_{\pi,typ}^{min}\le 200\,\mev$ with NLO or better 
ChPT or any other demonstrably controlled functional mass dependence

\item[\soso] $M_{\pi,typ}^{min}\le 350\,\mev$ and reliable estimate
of extrapolation error

\item[\bad] $M_{\pi,typ}^{min}> 350\,\mev$

\end{myitemizei}

\item {\bf continuum extrapolation}

\begin{myitemizei}

\item[\good] 3 or more lattice spacings with
at least one $a < 0.08\,\fm$ and controlled scaling

\item[\soso] 2 or more lattice spacings with one $a\lsim 0.1\,\fm$

\item[\bad] a single lattice spacing or all $a> 0.1\,\fm$

\end{myitemizei}

\item {\bf finite volume}

\begin{myitemizei}

\item[\good] $LM_\pi^{min}\ge 4$ (and numerical 
volume scaling study)

\item[\soso] $3 < LM_\pi^{min}\le 4$ and well motived analytical corrections

\item[\bad] $LM_\pi^{min}\le 3$ or $3 < LM_\pi^{min}\le 4$
and no quantification of finite-volume effects

\end{myitemizei}

where $M_\pi^{min}$ is the mass of the lightest pion contributing to
finite-volume effects.

\end{myitemize}

\subsection{Averaging of lattice results}
\label{sec:avg}

Now that results for various quantities of phenomenological interest are
emerging from lattice calculations in which most effects are realistically
taken into account, it is important to set forth objective, quantity
independent averaging procedures. In particular, that means taking literally
the statistical and systematic error estimates provided by the authors in
a refereed publication. It also means only considering calculations
in which all relevant sources of systematic uncertainty have been accounted
for. Since we are still in the early days of realistic lattice calculations,
this rule might have to be bent slightly at first to include results which are
close to reaching this goal.

The averaging procedure which I adopt is the following. Given a list of
results which satisfy the basic criteria described above, I perform their
weighted average, with an inverse weight obtained by adding the statistical
and systematic covariance matrices in quadrature. To determine the statistcal
error on the average, I construct a $\chi^2$ with only the statistical
correlation matrix and perform a standard $\Delta\chi^2$ analysis. For the
systematic error, since one does not generically expect them to compensate
from one calculation to the next, I take the smallest total systematic
uncertainty amongst those obtained in the most complete calculations. In cases
where either statistical or systematic errors are not symmetric, I symmetrize
them.

There will be some statistical correlations between results obtained from the
same set or from subsets of a given ensemble of gauge configurations. There
will also be some correlations in the systematic errors of calculations which
make use of similar methods. However, such correlations have not yet been
analyzed in any detail and I choose to neglect them here when computing world
averages. For computing an average's statistical error, though, I keep only
the statistical error of the calculation, amongst those performed on a same
set of configurations, that makes use of the largest fraction of these
configurations.  Correlations may be added more systematically later, once
they are better understood.

In situations where some results have significantly smaller
systematic uncertainties, for reasons which are not fully understood, one can
provide an average with and without those results.

\section{$|V_{us}|$ from experiment and the lattice}

A precise determination of the magnitude of the CKM matrix element $V_{us}$
allows for a precision test of CKM unitarity as well as of quark-lepton
universality and provides constraints on new physics, through:
\be
 \frac{G_q^2}{G_\mu^2} \Big[|V_{ud}|^2+|V_{us}|^2+|V_{ub}|^2\Big]=1+
O\left(\frac{M_W^2}{\Lambda_{NP}^2}\right)
\label{eq:unituniv}
\ ,\ee
where $(V_{ud},V_{us},V_{ub})$ forms the first row of the CKM matrix and where
$G_q$ is the Fermi constant as measured in quark decays, whereas $G_\mu$ is
the same constant as determined in muon decays. Eq.~(\ref{eq:unituniv})
accounts for the fact that what is actually measured are not the CKM matrix
elements, $V_{qq'}$, but $(G_q^2/G_\mu^2)\times|V_{qq'}|$. The large amounts
of new experimental results from BNL-E685, KLOE, KTEV, ISTRA+ and NA48 provide
the opportunity for testing this aspect of the standard model with
unprecedented accuracy.

The current situation on the measurement of the relevant CKM matrix elements
is:
\begin{myitemizeii}
\item
$|V_{ud}|=0.97425(22)\; {[0.02\%]}$ from nuclear $\beta$
decays~\cite{Hardy:2008gy}
\item
$|V_{us}|=0.2246(12)\; {[0.5\%]}$ from $K_{\ell 3}$ decays~\cite{Antonelli:2008jg}
\item
$|V_{us}/V_{ud}|=0.2321(15)\; {[0.6\%]}$ from $K_{\ell 2}$ decays~\cite{Antonelli:2008jg}
\item
$|V_{ub}|=3.87(47)\cdot 10^{-3}\; {[12\%]}$ from exclusive and inclusive $b\to
  u\ell\nu$ decays~\cite{CKMfitter09}
\end{myitemizeii}
where a factor of $(G_q/G_\mu)$ is implicit, as per \eq{eq:unituniv}, and
where the percentages in square brackets indicate, for convenience,
the relative error of the measurement. 

The Flavianet Kaon Working Group combined the first three measurement to
squeeze out a little additional precision on
$|V_{us}|$~\cite{Antonelli:2008jg}. I have updated their analysis here to take
into account the new result for $|V_{ub}|$~\cite{Hardy:2008gy}:
\begin{myitemizeii}
\item
$|V_{ud}|=0.97425(22)\; {[0.02\%]}$, which implies the following contribution
to the uncertainty in \eq{eq:unituniv}: $\delta |V_{ud}|^2= 4.3\cdot 10^{-4}$,
\item
$|V_{us}|=0.2252(9)\; {[0.4\%]}$, which implies the following contribution
to the uncertainty in \eq{eq:unituniv}: $\delta |V_{us}|^2= 4.2\cdot 10^{-4}$,
\item
and the contribution from $V_{ub}$ to \eq{eq:unituniv}, $|V_{ub}|^2\simeq
1.5\cdot 10^{-5}$, is so small that its error bar is irrelevant.
\end{myitemizeii}
At the time of the conference, $|V_{us}|$ was no longer the dominant source of
uncertainty in \eq{eq:unituniv}. However, the new result for
$|V_{ud}|$~\cite{Hardy:2008gy} makes it a dead heat. Combining these
results yields:
\be
\frac{G_q^2}{G_\mu^2}
\Big[|V_{ud}|^2+|V_{us}|^2+|V_{ub}|^2\Big]=0.9999(6)\;{[0.06\%]}
\ .
\ee
This result is fully consistent with the standard model. However, within one
standard deviation, new physics at a scale $\Lambda_{NP}\gsim 3\,\tev$ cannot
be excluded and within three standard deviations, this scale drops down to
$\Lambda_{NP}\gsim 2\,\tev$.

\begin{table}[t]
\begin{center}
\begin{tabular}{l@{\hspace{0mm}}c@{\hspace{1mm}}l@{\hspace{1mm}}c@{\hspace{3mm}}c@{\hspace{1mm}}c@{\hspace{1mm}}c@{\hspace{1mm}}}
\hline\hline
Ref.	&$N_f$&action		&$a[\fm]$
& \hspace{-2.mm}\begin{tabular}{c}$LM_\pi^{min}$\\[-0mm]
typ/val\end{tabular} & \hspace{-2.mm}\begin{tabular}{c}$M_\pi^{min}[\mev]$\\[-0mm]
typ/val\end{tabular} &
	\multicolumn{1}{c}{$F_K/F_\pi$} 
\\
\hline
\multicolumn{6}{l}{\hspace{-2.mm}PDG'08~\cite{Amsler:2008zzb}}
&1.193(6)
\\
\hline

\hspace{-2.mm}{ ETM'08~\cite{Blossier:2008dj}}
& 2 & tmQCD & $0.07,0.09,0.10[F_\pi]$ & 3.6/3.6 
& $260/260$ & {\em  1.196(13)(7)(8)}\\[-0mm]
\hspace{-2.mm}\begin{tabular}{l}NPLQCD'06\\[-2mm]\cite{Beane:2006kx}
\end{tabular}
& 2+1& {\hspace{-2.mm}\begin{tabular}{l}$\rm DWF/$
\\[-2mm]
$\rm 
KS_{\rm MILC}$	\end{tabular} }
	&   $0.13[r_0]$ &5.1/3.5& 
$420/290$&$1.218(2)^{+11}_{-24}$\\[-0mm]
\hspace{-2.mm}\begin{tabular}{l}MILC'04-07\\[-2mm]
\cite{Aubin:2004fs,Bernard:2007ps}\end{tabular}
& 2+1 & ${\rm KS_{\rm MILC}^{\rm AsqTad}}$ & 
\hspace{-2.mm}\begin{tabular}{c}$0.06,0.09,0.12,$\\[-0mm]
$0.15[F_\pi]$ \end{tabular}
& 5.3/4.2 &
	$300/240$&$1.197(3)^{+6}_{-13}$ 
\\[-0mm]
{\hspace{-2.mm}\begin{tabular}{l}HPQCD/'07\\[-2mm]
	UKQCD~\cite{Follana:2007uv}
	\end{tabular} } &2+1& $\rm KS^{\rm HISQ}_{\rm MILC}$ & $0.09,0.12,
0.15[\Upsilon]$ & 4.8/4.1 &$360/310$& 1.189(2)(7)\\ [-0mm]
{\hspace{-1.6mm}\begin{tabular}{l} RBC/'08
\\[-2mm]UKQCD~\cite{Allton:2008pn}
\end{tabular} }
&2+1& DWF		&    $0.11[\Omega]$ & 4.1/3.4	&	$290/240$ &1.205(18)(62) 
\\[-0mm]
ALV'08 & 2+1& {\hspace{-2.mm}\begin{tabular}{l}$\rm DWF/$
\\[-2mm]
$\rm 
KS_{\rm MILC}$	\end{tabular} }
	&   $0.09,0.12[\Upsilon/F_\pi]$ & 5.3/4.2 &$300/240$ & \em  
1.191(16)(17)\\[-0mm]
\hspace{-2mm}PACS-CS'08\cite{Aoki:2008sm}
& 2+1& NP-SW &    $0.09[\Omega]$ & 2.3/2.3	&	$160/160$&{1.189(20)} 
\\[-0mm]
{BMW'08}
 & 2+1 & SW & \hspace{-2.mm}\begin{tabular}{c}$0.065,0.085,$\\[-0mm]
$0.125[\Xi]$ \end{tabular} & $4/4$ & $190/190$ & \em  1.19(1)(1)
\\
\hline
\hline
\end{tabular}
\end{center}
\caption{\label{tab:FKFpi} \em Parameters of the simulations used by various
  collaborations for calculating $F_K/F_\pi$, together with their result for
  that quantity (results in italic were presented at this conference). The
  column $N_f$ indicates the number of sea quark flavors considered. The
  symbols in brackets in the $a$[fm] column indicate the quantity used to set
  the scale. Also given are the masses of the lightest valence pion simulated,
  as well as the ``typical'' lightest pion mass defined in
  footnote~\protect\ref{foot:mpityp}.
\vspace{-0.3cm}}
\end{table}

\subsection{$|V_{us}/V_{ud}|$ from $K,\,\pi\to\mu\bar\nu$}

In 2004, Marciano pointed out a window of opportunity for determining
$|V_{us}/V_{ud}|$ from the ratio of leptonic decay rates
$\Gamma(K\to\mu\bar\nu(\gamma))/\Gamma(\pi\to\mu\bar\nu(\gamma))$~\cite{Marciano:2004uf}. Calculating
$O(\alpha)$ radiative corrections to this ratio, he obtained (see update in
\cite{Amsler:2008zzb}):
\be
\frac{|V_{us}|}{|V_{ud}|}\frac{F_K}{F_\pi} =
0.2757(7)\;
  [0.25\%]
\ .\ee
Thus, a precise lattice calculation of $F_K/F_\pi$ will allow a high precision
determination of $|V_{us}/V_{ud}|$. One needs to determine $F_K/F_\pi$ to:
\begin{itemize}
\item
0.5\% to match the uncertainty on $|V_{us}|$
obtained in $K\to\pi\ell\nu$ decays,
\item
0.25\% to match the experimental uncertainty in
$\Gamma(K\to\mu\bar\nu(\gamma))/\Gamma(\pi\to\mu\bar\nu(\gamma))$.
\end{itemize}
$F_K/F_\pi$ is an $SU(3)$-flavor breaking effect, i.e.
\be
F_K/F_\pi = 1 + O\left(\frac{M_K^2 - M_\pi^2}{\Lambda_\chi^2}
\right)
\ee
and it is the deviation from unity that we are actually calculating, which
makes the target accuracies a little less forbidding.

In \tab{tab:FKFpi} I summarize the parameters and results of all unquenched
lattice calculation of $F_K/F_\pi$. The corresponding consumer report is given
in \tab{tab:FKFpiflavia}.

\begin{table}[t]
\begin{center}
\begin{tabular}{l@{\hspace{10mm}}c@{\hspace{10mm}}c@{\hspace{10mm}}c@{\hspace{10mm}}c@{\hspace{10mm}}c@{\hspace{10mm}}c@{\hspace{14mm}}}
\hline
\hline\\[0.4cm]
Ref. & \begin{rotate}{40}{\footnotesize publication}\end{rotate}
& \begin{rotate}{40}{\footnotesize action, unit.}\end{rotate}
& \begin{rotate}{40}{\footnotesize $N_f$}\end{rotate}
& \begin{rotate}{40}{\footnotesize mass extrap}\end{rotate}
& \begin{rotate}{40}{\footnotesize $a\to 0$}\end{rotate}
& \begin{rotate}{40}{\footnotesize finite volume}\end{rotate}
\\
\hline
{ETM'08~\cite{Blossier:2008dj}} & \bad & \good & \soso & \soso & \good & \soso \\
{NPLQCD'06~\cite{Beane:2006kx}} & \good & \soso & \good &  \bad & \bad & \good \\
{MILC'04-07~\cite{Aubin:2004fs,Bernard:2007ps}} 
& \good & \soso & \good & \soso  & \good & \good \\
{HP/UKQCD'07~\cite{Follana:2007uv}} 
& \good & \soso & \good & \bad & \good & \good \\
{RBC/UKQCD'08~\cite{Allton:2008pn}} & \good & \good & \good & \soso & \bad & \good\\
{ALV'08~\cite{Aubin:2008ie}} & \bad & \soso & \good & \soso & \soso & \good\\
{PACS-CS'08~\cite{Aoki:2008sm}} & \soso & \good & \good & \good & \bad & \bad\\
{BMW'08} & \bad & \good & \good & \good & \good & \good\\
\hline
\hline
\end{tabular}
\caption{\label{tab:FKFpiflavia} \em Starring of the simulations used to
  obtain $F_K/F_\pi$, according to the criteria put forth in
  \protect\sec{sec:stars}. \vspace{-0.3cm}}
\end{center}
\end{table}

Of all these calculations, the most advanced is that of
MILC~\cite{Aubin:2004fs,Bernard:2007ps}, but the calculation of the BMW
collaboration, presented at this conference by S. D\"urr, should rival it once
completed. The calculation of PACS-CS~\cite{Aoki:2008sm}, performed very close
to the physical up and down quark mass holds great promise. However, as it
stands, it is missing a continuum extrapolation and may also suffer from
significant finite-volume errors.

\begin{figure}[t]
\begin{center}
\psfig{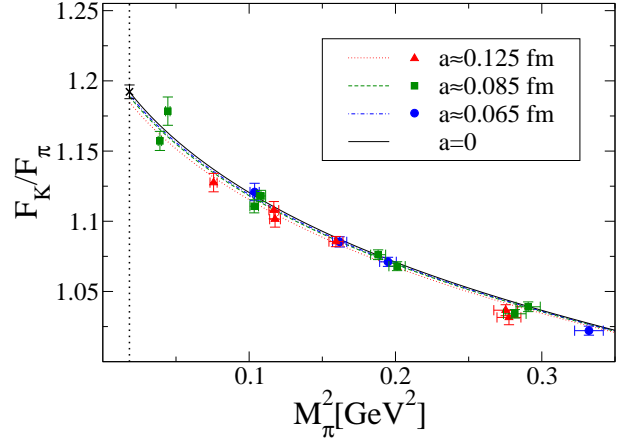}
\caption{\label{fig:BMWFKFpi} \em $F_K/F_\pi$ versus $M_\pi^2$ in physical
  units, as presented by BMW at this conference. The different symbols
  correspond to different lattice spacings, as indicated in the plot
  legend. The curves represent the result of a combined chiral and continuum
  extrapolation fit for each lattice spacing, as well as in the continuum
  limit. The results have already been interpolated in strange quark mass to
  the physical value. The particular fit shown corresponds to a NLO, $SU(3)$
  ChPT fit with $M_\pi< 470\,\mev$ and $a^2$ discretization errors.
\vspace{-0.5cm}}
\end{center}
\end{figure}

To illustrate lattice results for $F_K/F_\pi$, let me briefly present those of
BMW. The calculations are performed in volumes up to 4~fm, with pions as light
as 190~MeV and lattice spacings down to 0.065~fm. The parameters of the
calculation are summarized in \tab{tab:FKFpi}, and details of the ensembles
can be found in \cite{Durr:2008zz}. The results are plotted in
\fig{fig:BMWFKFpi}, as a function of $M_\pi^2$ in physical units, with the
scale set by the $\Xi$ mass as in \cite{Durr:2008zz}. The plot shows the
extrapolation of the results for $F_K/F_\pi$ in $M_\pi^2$ to the physical
point. A large variety of functional forms have been tried, ranging from NLO
$SU(2)$ ChPT to polynomial expansions.  Three different cuts on pion mass have
been made: $M_\pi< 420\,\mev$, 470~MeV and 600~MeV. The continuum and mass
extrapolations are combined, by allowing for the parameters of the functional
mass dependence to acquire $a^2$ or $a$ corrections. Finite-volume effects are
subtracted at two-loops in ChPT, using the results of
\cite{Colangelo:2005gd}. The procedure for estimating statistical and
systematic uncertainties is very similar to that in \cite{Durr:2008zz}. It
should be noted that the shift in $F_K/F_\pi$ from the lightest pion mass to
the physical point is less than 2\%. The preliminary result is given in
\tab{tab:FKFpi}.

\begin{figure}[t]
\begin{center}
\psfig{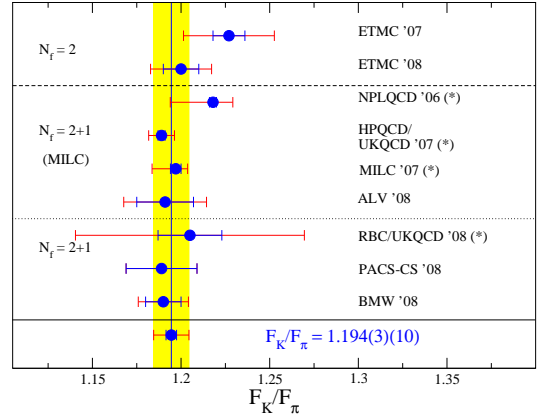}
\caption{\label{fig:FKFpi} \em Summary of unquenched lattice results for
  $F_K/F_\pi$, together with my average. The latter is obtained as described
  in \protect\sec{sec:avg} and in the text. The results marked with a
  ``(*)'' are those included in the average. The smallest error bar on each
  point is the statistical error and the larger one, the statistical and
  systematic errors combined in quadrature. The references are as in
  \protect\tab{tab:FKFpi}.\vspace{-0.5cm} }
\end{center}
\end{figure}

Unquenched, lattice results for $F_K/F_\pi$ are summarized in \fig{fig:FKFpi},
where my average for this quantity, obtained as explained in \sec{sec:avg}, is
also given. This average includes only the published $N_f=2+1$
results~\cite{Beane:2006kx,Aubin:2004fs,Bernard:2007ps,Follana:2007uv,
  Allton:2008pn} in which many systematic uncertainties have been
estimated. The systematic error is taken from \cite{Bernard:2007ps}. The total
uncertainty on this quantity is $\delta (F_K/F_\pi)^{lat}=0.8\%$. This
corresponds to an uncertainty of $\delta (F_K/F_\pi-1)^{lat}\simeq 5\%$ on the
calculated $SU(3)$-flavor breaking effect, which is much better than the
accuracy obtained on the $SU(3)$-flavor breaking in the form factor for
$K\to\pi\ell\nu$, $\delta f_+(0)\simeq 15\%$. Nevertheless, this uncertainty
still leads to a larger theory error in the determination of $|V_{us}|$,
i.e. 0.8\% vs 0.5\%. Since $F_K/F_\pi$ is a straightfoward quantity to
calculate, one may expect steady improvements in its lattice determination,
especially in light of the recent progress by PACS-CS~\cite{Aoki:2008sm}.

\subsection{$|V_{us}|$ from $K\to\pi\ell\nu$}

$K\to\pi\ell\nu$ decays provide an alternative way to determine
$|V_{us}|$. This measurement requires the theoretical calculation of the
vector form factor, $f_+(q^2)$, defined through:
\be
\la\pi^+(p')|\bar u\gamma_\mu s|\bar K^0(p)\ra
=(p+p'-q\frac{M_K^2-M_\pi^2}{q^2})_\mu f_+(q^2)+
q_\mu\frac{M_K^2-M_\pi^2}{q^2}f_0(q^2)
\ ,\ee
with $q=p-p'$. The best precision is currently obtained by measuring the form
factor shape in experiment and extracting, from the total
rate~\cite{Antonelli:2008jg},
\be
|V_{us}| \times f_+(0) = 0.21664(48)\; {[0.22\%]}
\ .
\ee
The experimental error is of similar size as in the ratio of leptonic kaon to
pion decay rates. To fully exploit the experimental results requires a
determination of $f_+(0)$ to 0.22\%.

To the extent that it is applicable here, the theoretical framework for
attacking this problem is $SU(3)$ ChPT
\cite{Gasser:1984gg,Leutwyler:1984je}. The chiral expansion for $f_+(0)$ is
given by:
\be
f_+(0)=1+f_2+f_4+\cdots
\ ,\ee
where the Ademollo-Gatto theorem~\cite{Ademollo:1964sr} and ChPT yield:
\be
f_2 = 
O\left(\frac{(M_K^2-M_\pi^2)^2}{M_K^2\Lambda_\chi^2}\right)
=-0.023
\ .
\ee
Thus, there are no contributions for the $O(p^4)$ LECs and this NLO
contribution is fully determined by $M_K$, $M_\pi$ and $F_\pi$.

This means that a sub-percent level determination of $f_+(0)$ requires a
calculation of NNLO and higher corrections, since
\be
\Delta f\equiv f_+(0)-1-f_2=O\left(\frac{(M_K^2-M_\pi^2)^2}{\Lambda_\chi^4}
\right)\sim 3\%
\ee
is comparable in size to $f_2$. To fully exploit the experimental accuary
``only'' requires an accuracy of 7\% in the calculation of $\Delta f$.

What is known about $f_4$ and more generally $\Delta f$? The NNLO chiral logs
have been computed~\cite{Post:2001si,Bijnens:2003uy}, and they require
$O(p^6)$ LECs for renormalization. Estimates have been made of these LECs
\cite{Bijnens:2003uy,Jamin:2004re,Cirigliano:2005xn,Portoles:2007gb} and in
\cite{Bijnens:2003uy} it is shown that they can be determined from the slope
and curvature of $f_+(q^2)$. The reference value for $\Delta f$ is still taken
to be the quark model result, $\Delta f=-0.016(8)$~\cite{Leutwyler:1984je}.

In \tab{tab:Fp0}, I summarize the parameters and results of all unquenched
lattice calculations of $f_+(0)$. The corresponding consumer report is given
in \tab{tab:Fp0flavia}.

\begin{table}
\begin{center}
\begin{tabular}{lllcccc}
\hline
\hline
Ref. & $N_f$ & action & $a[\fm]$ & $L[\fm]$ & 
\begin{tabular}{c}$M_\pi^{min}[\mev]$\\[-0mm]
typ/val\end{tabular} & $f_+(0)$\\
\hline
{JLQCD'05~\cite{Tsutsui:2005cj}} & 2 & NP SW & 0.09 & 1.8 & 550/550 & 0.967(6)\\
{RBC'06~\cite{Dawson:2006qc}} & 2 & DWF & 0.12 & 2.5 & 490/490 & 0.968(9)(6)\\
{ETM'08~\cite{Simula:CKM08}} & 2 & tmQCD & 0.11 & 2.7 & 260/260 & 0.957(5)\\
{FNAL/MILC'04~\cite{Okamoto:2004df}} & 2+1 & KS+Wil &  &  & & 0.962(6)(9)\\
{RBC/UKQCD'07\cite{Boyle:2007qe}} & 2+1 & DWF & 0.11 & 1.8, 2.8 & 290/240 & 0.9644(33)(34)(14)\\
\hline
\hline
\end{tabular}
\caption{\label{tab:Fp0} \em Parameters of the simulations used by various
  collaborations for calculating $f_+(0)$, together with their result for that
  quantity. The description of the columns can be inferred from the one
  given in \protect\tab{tab:FKFpi}.
\vspace{-0.3cm}}
\end{center}
\end{table}

\begin{table}
\begin{center}
\begin{tabular}{l@{\hspace{10mm}}c@{\hspace{10mm}}c@{\hspace{10mm}}c@{\hspace{10mm}}c@{\hspace{10mm}}c@{\hspace{10mm}}c@{\hspace{14mm}}}
\hline
\hline\\[0.4cm]
Ref. & \begin{rotate}{40}{\footnotesize publication}\end{rotate}
& \begin{rotate}{40}{\footnotesize action, unit.}\end{rotate}
& \begin{rotate}{40}{\footnotesize $N_f$}\end{rotate}
& \begin{rotate}{40}{\footnotesize mass extrap}\end{rotate}
& \begin{rotate}{40}{\footnotesize $a\to 0$}\end{rotate}
& \begin{rotate}{40}{\footnotesize finite volume}\end{rotate}
\\
\hline
{JLQCD'05~\cite{Tsutsui:2005cj}} & \bad & \good & \soso &  \bad & \bad & \good \\
{RBC'06~\cite{Dawson:2006qc}} & \good & \good & \soso & \bad  & \bad & \good \\
{ETM'08~\cite{Simula:CKM08}} & \bad & \good & \soso & \soso & \bad & \soso \\
{FNAL/MILC'04~\cite{Okamoto:2004df}} & \bad & \soso & \good & & &  \\
{RBC/UKQCD'07\cite{Boyle:2007qe}} & \good & \good & \good & \soso & \bad & \good\\
\hline
\hline
\end{tabular}
\caption{\label{tab:Fp0flavia} \em Starring of the simulations used to
  obtain $f_+(0)$, according to the criteria put forth in
  \protect\sec{sec:stars}. \vspace{-0.3cm}}
\end{center}
\end{table}

\begin{figure}[t]
\begin{center}
\psfig{file=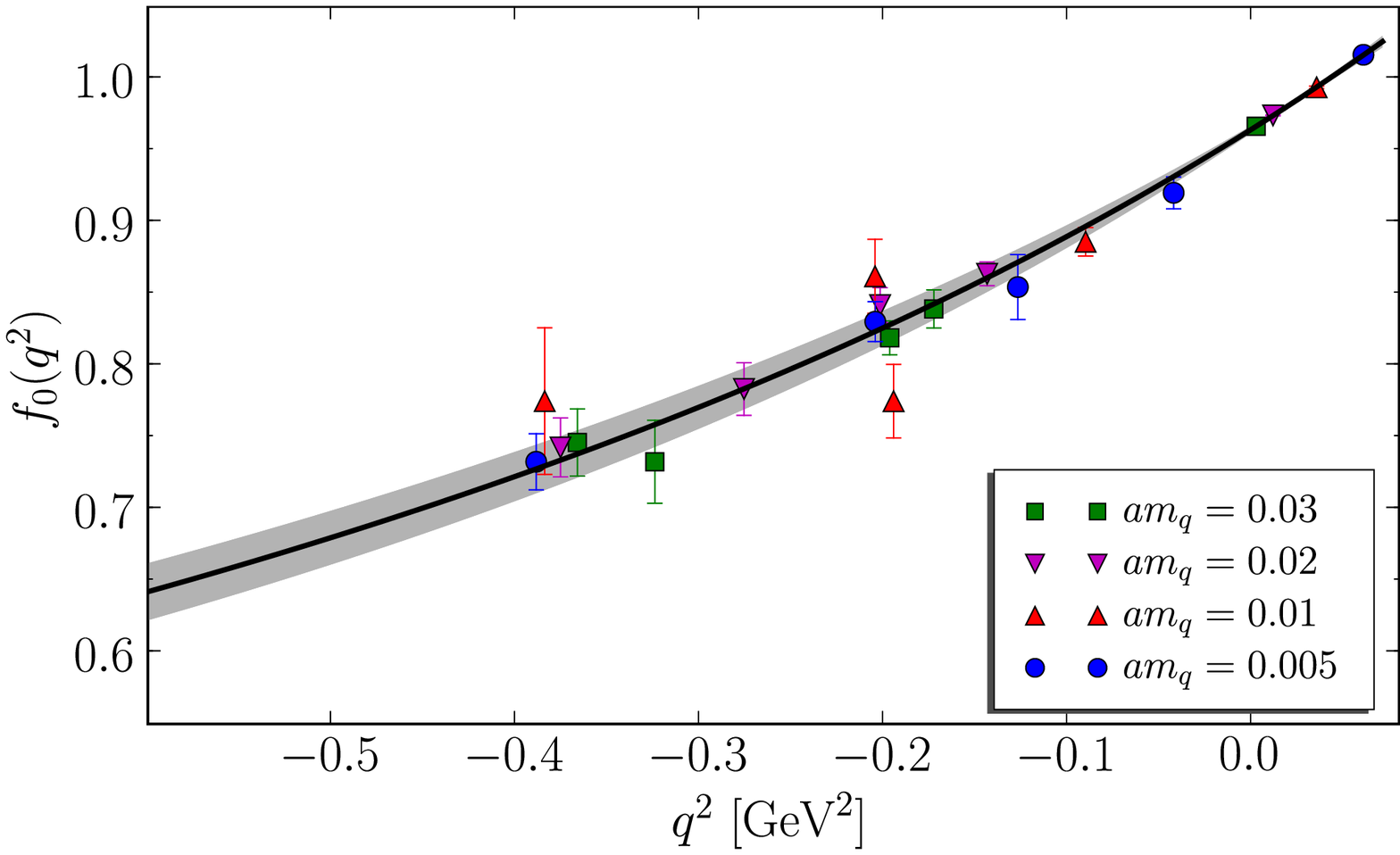,width=7cm}
\psfig{file=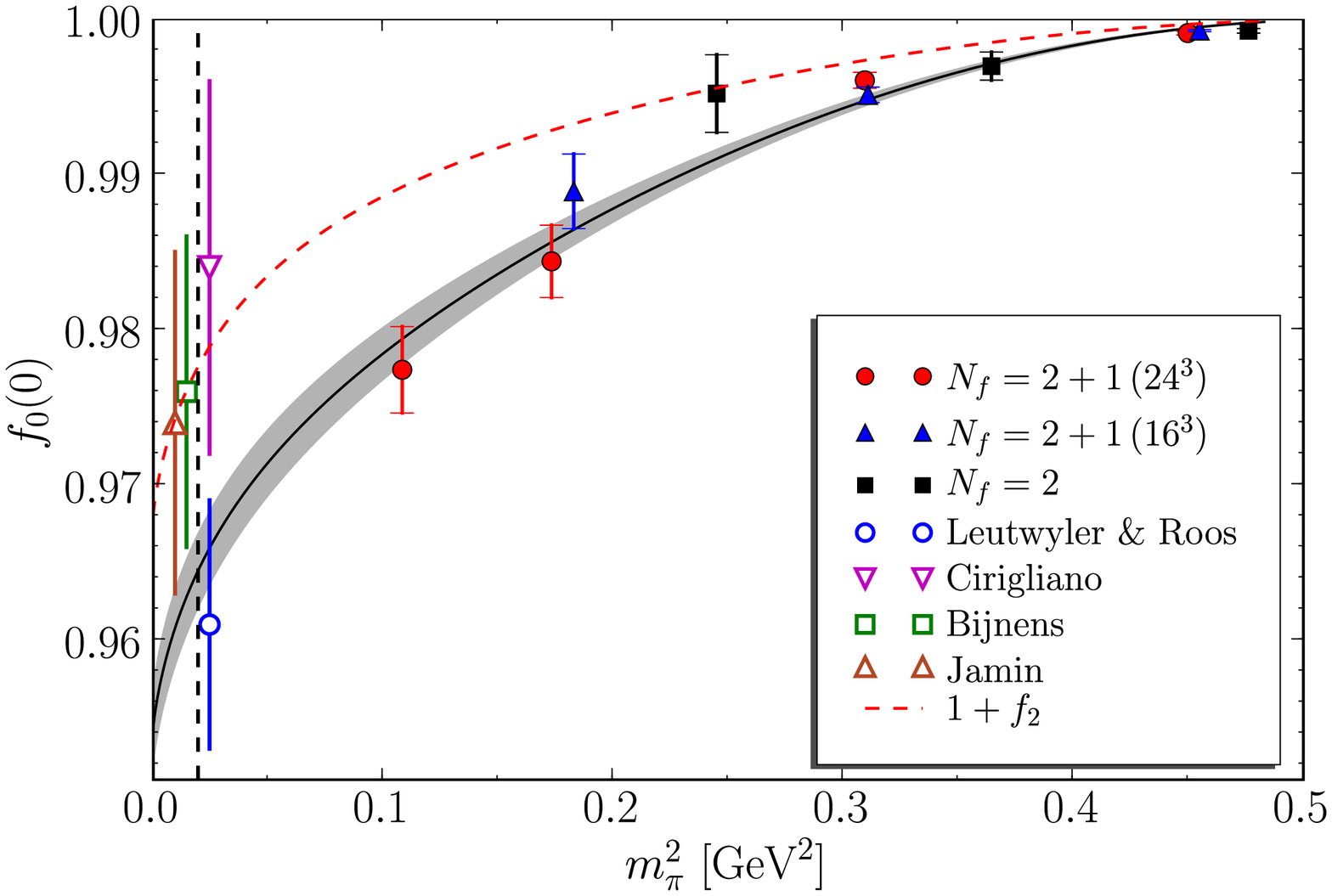,width=6.5cm}
\caption{\label{fig:RBCf0} \em Combined $q^2$ and chiral fit of $f_0(q^2)$ to
  \protect\eq{eq:f0q2chi} by RBC/UKQCD~'08~\cite{Boyle:2007qe}. The left panel
  displays the lattice values of $f_0(q^2)$ vs $q^2$, together with the fit
  curve obtained at the physical values of $M_\pi$ and $M_K$. The lattice
  points were shifted in pion and kaon mass at fixed $q^2$ using the fit
  result. The right panel displays the extrapolation of $f_+(0)=f_0(0)$ in
  $M_\pi^2$ to physical pion mass.\vspace{-0.5cm}}
\end{center}
\end{figure}

The lattice methodology for the calculation of $f_+(0)-1 $ was set forth in
\cite{Becirevic:2004ya}. It consists of three main steps:
\begin{enumerate}

\item Use a double ratio of three-point functions to obtain:
\be
f_0(q^2_{max}) = \frac{2\sqrt{M_KM_\pi}}{M_K+M_\pi}\frac{\la\pi\vert 
V_0\vert K\ra\la K\vert V_0\vert\pi\ra}{\la\pi\vert 
V_0\vert \pi\ra\la K\vert V_0\vert K\ra}
\ .\ee
This yields a determination of $f_0(q^2_{max})$ with a statistical error less
than about $0.1\%$!

\item
Compute $f_0(q^2)$ at various $q^2$ and use an ansatz to interpolate and get
$f_+(0)=f_0(0)$. 

\item
Interpolate/extrapolate in light quark mass to the physical mass point.
\end{enumerate}
RBC/UKQCD~\cite{Boyle:2007qe} actually combine steps 2 and 3, using the
functional form:
\be
f_0(q^2;M_K,M_\pi)=\frac{1+f_2(M_K,M_\pi)+(M_K^2-M_\pi^2)^2({A_0}+{A_1}
(M_K^2+M_\pi^2))}{1-
q^2/({M_0}+{M_1}(M_K^2+M_\pi^2))^2}
\ ,
\label{eq:f0q2chi}\ee
where $A_0$, $A_1$, $M_0$ and $M_1$ are parameters and where a polynomial
ansatz was made for NNLO terms. This combined fit is shown in the two panels
of \fig{fig:RBCf0}.
Their results fit $1+f_2(M_K,M_\pi)+\mathrm{NNLO}$ well, though their fits do
not take correlations into account. The claim that they are sensitive
to NNLO effects seems to be justified. Moreover, the extrapolated result is
only two standard deviations below the result obtained at their lightest pion
mass and the claimed error on $f_+(0)-1$ is a rather conservative 14\%. The
caveats are that $m_s$ is approximately 15\% too high and the calculations
were performed at a single, rather coarse lattice spacing of
$a=0.114(2)\,\fm$, meaning that discretizations errors can only be guessed.
Nevertheless, this is the first convincing lattice calculation of $f_+(0)-1$.

\begin{figure}[t]
\begin{center}
\psfig{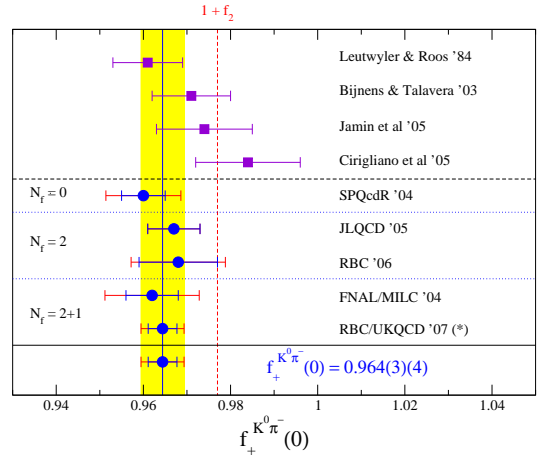}
\caption{\label{fig:fp0summary} \em Summary of lattice results for $f_+(0)$,
  together with the results obtained in various models. Also given is my
  average of the unquenched lattice resutls. The latter is obtained as
  described in \protect\sec{sec:avg} and in the text. The results marked with
  a ``(*)'' are those included in the average. The smallest error bar on each
  point is the statistical error and the larger one, the statistical and
  systematic errors combined in quadrature. The references for the lattice
  works are as in \protect\tab{tab:Fp0} with, in addition,
  SPQcdR~'04~\cite{Becirevic:2004ya}. The others are
  Leutwyler and Roos~'84~\cite{Leutwyler:1984je}, Bijnens and
  Talavera~'03~\cite{Bijnens:2003uy}, Jamin et al~'05~\cite{Jamin:2004re},
  Cirigliano et al~'05~\cite{Cirigliano:2005xn}.\vspace{-0.5cm}}
\end{center}
\end{figure}

Lattice and non-lattice results for $f_+(0)$ are summarized in
\fig{fig:fp0summary}, together with the ``average'' which I obtain by copying
the result of~\cite{Boyle:2007qe}.
The total uncertainty on $f_+(0)$ is $\delta f_+(0)^{lat}=0.5\%$. This means
that $K\to\pi\ell\nu$ decays still give the best accuracy for
$|V_{us}|$. Moreover, one can anticipate that the current error, $\delta
(f_+(0)-1)^{lat}=14\%$ will be reduced thanks to the use of: stochastic
sources, as used in~\cite{Simula:2007fa,Boyle:2008yd}; partially twisted
boundary conditions~\cite{Bedaque:2004ax,Flynn:2005in}, applied to form
factors
in~\cite{Simula:2007fa,Boyle:2008yd,Guadagnoli:2005be,Boyle:2007wg,Boyle:2008hd},
which enable to determine $f_+(q^2)$ directly at
$q^2=0$~\cite{Boyle:2007wg,Boyle:2008yd}.

\section{$K\to\pi\pi$ decays on the lattice}

The phenomenology of $K\to\pi\pi$ decays is extremely rich, and has been
highly instrumental in developing the standard model. In the isospin limit,
the amplitudes for these decays can be decomposed in terms of amplitudes
$A_Ie^{i\delta_I}$, $I=0,2$, where $I$ is the isospin of the final two-pion
state and $\delta_I$ is the strong scattering phase in that channel. CP
violation implies that $A_I^*\ne A_I$. CP violation occurs in two ways in
$K_L$ decays. $K_L$ is mostly CP odd, and decays predominently into three
pions. But it has a small CP even component, through which it can also decay
into two pions. This process is known as indirect CP violation, and is
parametrized by~\cite{deRafael:1995zv,Anikeev:2001rk,Buras:2008nn}:
\be
\epsilon= \frac{T[K_L\to(\pi\pi)_{I=0}]}{T[K_S\to(\pi\pi)_{I=0}]}
\simeq e^{i\phi_\epsilon}\sin\phi_\epsilon\left(
\frac{\im M_{12}}{\Delta M_K}+\xi\right)
\label{eq:epsK}
\ ,\ee
with $\xi=\im A_0/\re A_0$, $\Delta M_K \equiv M_{K_L}-M_{K_S}$ and $M_{12}$
to be defined below. $K_L$ decays can violate CP through another channel, by
having its CP odd component decay directly into two pions. This process is
known as direct CP violation, and is parametrized by~\cite{deRafael:1995zv}:
\bea
\epsilon'&=&\frac{1}{\sqrt2}\frac{T[K_s\to(\pi\pi)_{I=2}]}{T[K_S\to(\pi\pi)_{I=0}]}\left[\frac{T[K_L\to(\pi\pi)_{I=2}]}{T[K_S\to(\pi\pi)_{I=2}]}-\frac{T[K_L\to(\pi\pi)_{I=0}]}{T[K_S\to(\pi\pi)_{I=0}]}\right]\nonumber\\ &
\simeq
&\frac{1}{\sqrt2}e^{i(\pi/2+\delta_2-\delta_0)}\frac{\re A_2}{\re A_0}
\left[\frac{\im A_2}{\re A_2}-\frac{\im A_0}{\re A_0}\right]\ .
\eea

Experimentally a lot is known about these different
processes~\cite{Amsler:2008zzb}. The $K_L$-$K_S$ mass difference is measured to
high precision, i.e. $\Delta M_K = (3.483\pm 0.006)\times
10^{-12}\,\mev\;{[0.2\%]}$. $K\to\pi\pi$ decays exhibit a strong enhancement
of the $I=0$ channel over the $I=2$ channel, $|A_0/A_2|\simeq 22.2$, known as
the $\Delta I=1/2$ rule, which is still in need of an explanation after over
forty years. The parameter for indirect CP violation has also been measured to
high accuracy, $|\epsilon|=(2.229\pm 0.012)\cdot 10^{-3}\; {[0.5\%]}$, with a
phase $\phi_\epsilon=(43.5\pm 0.7)^o\; {[1.6\%]}$. And after an experimental
effort of nearly thirty years, direct CP violation was also measured, yielding
$\re(\epsilon'/\epsilon)=(1.65\pm 0.26)\cdot 10^{-3}\;{[16\%]}$.

\subsection{$K^0$-$\bar K^0$ mixing in the standard model and $B_K$}

$K^0$-$\bar K^0$ mixing is responsible for the $K_L$-$K_S$ mass difference as
well as for indirect CP violation in $K\to\pi\pi$. In the standard model, the
CP violating contribution occurs through a local
$\Delta S=2$, four-quark operator, once the heavy, standard model degrees of
freedom are integrated out. The corresponding amplitude is
$$
2M_KM_{12}^*=\la\bar K^0|\cH_\mathrm{eff}^{\Delta S=2}|K^0\ra=
C_1^\mathrm{SM}(\mu)\la\bar K^0|O_1(\mu)|K^0\ra
\ ,$$
where $C_1^\mathrm{SM}$ is a short-distance Wilson coefficient and where
\be
O_1=(\bar sd)_{V-A}(\bar sd)_{V-A}\qquad\mathrm{and}\qquad
\la\bar K^0|O_1(\mu)|K^0\ra=\frac{16}3
M_K^2F_K^2{B_K(\mu)}
\ .\ee

\begin{figure}
\begin{center}
\psfig{file=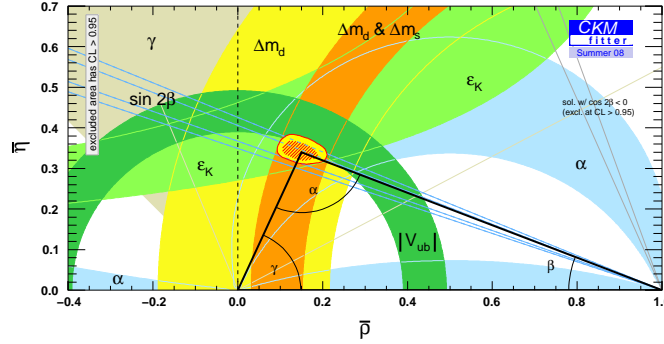,angle=0,width=10cm}
\caption{\label{fig:ckmfitter09} \em Constraints on the summit
  $(\bar\rho,\bar\eta)$ of the unitarity triangle from a global CKM
  fit~\protect\cite{CKMfitter09}.\vspace{-0.5cm}}
\end{center}
\end{figure}

In terms of theses quantities, a revised~\cite{Anikeev:2001rk,Buras:2008nn}
standard model analysis~\cite{Buchalla:1995vs} yields:
\be
|\epsilon|\simeq\kappa_\epsilon C_\epsilon\hat B_K
\left[\im(\lambda_t^{*2})
\eta_{tt}S_0(x_t)+
2\im(\lambda_t^*\lambda_c^*)\eta_{ct}S_0(x_c,x_t)+
\im(\lambda_c^{*2})\eta_{cc}S_0(x_c)\right]
\ ,
\label{eq:epsKSM}
\ee
where $C_\epsilon$ is determined by well measured quantities, $\hat
B_K=C_1^\mathrm{SM}(\mu)B_K(\mu)$ is the renormalization-group invariant
$B$-parameter, $\lambda_q\equiv V_{qd}V_{qs}^*$ and $\eta_{qq'}$, $S_0$ are
short-distance quantities. $\kappa_\epsilon$ parametrizes the corrections to
the standard analysis~\cite{Buchalla:1995vs}, which arise from
$\sqrt2\sin\phi_\epsilon$ and $\xi$ in \eq{eq:epsK} at leading order. This
approximation is not necessary and should not be used for precision tests of
the SM once $\xi$ is known better. A rough estimate yields
$\kappa_\epsilon=0.92(2)$~\cite{Buras:2008nn}, which implies an $8\pm2\%$
downward shift in the SM prediction for $\epsilon$.  From \eq{eq:epsKSM}, it
is clear that a measurement of $|\epsilon|$ and a determination of $B_K$
imposes contraints on $\im{\lambda_t^{*2}}$, $\im{\lambda_c^{*2}}$ and
$\im{\lambda_t^*\lambda_c^*}$, as shown in \fig{fig:ckmfitter09}.

\begin{figure}
\begin{center}
\psfig{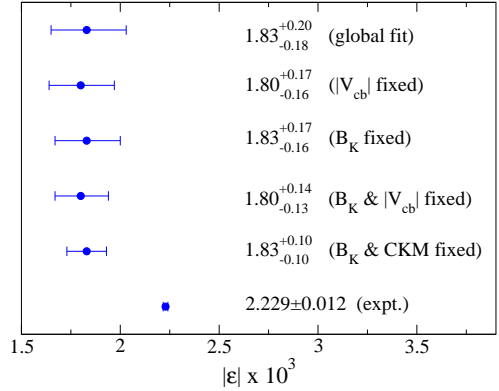}
\caption{\label{fig:epserrors} \em Standard model predictions for $|\epsilon|$,
  using results from global CKM fits by
  CKMfitter~\protect\cite{CKMfitter09,charles09}, compared to experiment. The
  various predictions are obtained by fixing different inputs to their central
  values, as described in the plot legend and in the text.\vspace{-0.5cm}}
\end{center}
\end{figure}

\begin{figure}
\begin{center}
\psfig{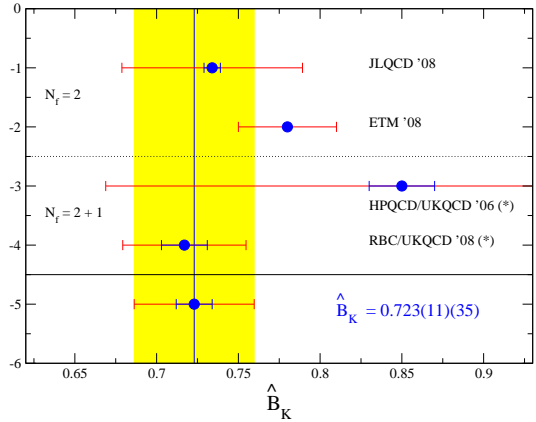}
\caption{\label{fig:BKsummary} \em Summary of unquenched lattice results for
  the renormalization group invariant $\hat B_K$, together with my
  average. The latter is obtained as described in \protect\sec{sec:avg} and in
  the text. The smallest error bar on each point is the statistical error and
  the larger one, the statistical and systematic errors combined in
  quadrature. The results marked with a ``(*)'' are those included in the
  average. The references are as in \protect\tab{tab:BKparam}.\vspace{-0.5cm}}
\end{center}
\end{figure}

Given how accurately $|\epsilon|$ is measured, one may wonder why the
constraint that it gives on the summit of the triangle is not any better. To
help answer this question, in \fig{fig:epserrors} I display SM predictions for
$|\epsilon|$ obtained in different ways. The starting point is a global CKM
fit using \eq{eq:epsKSM}, in which the experimental measurement for
$|\epsilon|$ is not included and where $\hat B_K=0.723(11)(35)\;{[5\%]}$ (from
\fig{fig:BKsummary}) and
$|V_{cb}|=0.04059(38)(58)\;{[1.7\%]}$~\cite{CKMfitter09}. The topmost
theoretical prediction for $|\epsilon|$ is obtained from this global fit,
allowing all quantities to fluctuate within their error
bars.~\footnote{Systematic errors due to theory are assumed to have a {\em
    gaussian} distribution, so that the deviation of the predicted from the
  measured $|\epsilon|$ can be counted in standard deviations.} The next
result is obtained by freezing $|V_{cb}|$ to its central value. The third
value results from fixing $B_K$ to its central value. The fourth is obtained
by freezing both $|V_{cb}|$ and $B_K$, and the fifth by fixing $B_K$ to its
central value and the four CKM parameters to their best global fit values. The
last is the experimental measurement quoted above.

As the second point indicates, a determination of $|V_{cb}|$ to infinite
accuracy only reduces the uncertainty on the prediction for $|\epsilon|$ from
10\% to 9\%. Significantly improving the accuracy on $B_K$ has a similar
effect, since the uncertainty on $|\epsilon|$ is also 9\% in that case.  The
fourth point indicates that the uncertainty coming from sources other than
$B_K$ and $|V_{cb}|$ is a little less than 8\%. It is only when the
uncertainties on $B_K$ and CKM parameters are assumed to be zero that the
error on the SM prediction for $|\epsilon|$ falls to 5\% . The
latter is due to perturbative uncertainties and to the error on
$\kappa_\epsilon$.

It is interesting to note that the SM prediction for $|\epsilon|$ is now about
18\% below the experimental value, down from what it was until recently. The
decrease is mainly due to the fact that the central value for $\hat B_K$ has
dropped by more than 15\% over the last decade (see discussion below and
caveats) and to the presence of the correction related to $\kappa_\epsilon$ in
\eq{eq:epsKSM}. This potential tension between theory and observation has led
the authors of \cite{Lunghi:2008aa,Buras:2008nn,Buras:2009pj} to investigate
new CP violating contributions to $\Delta F=2$ observables. For the moment, as
the top point in \fig{fig:epserrors} indicates, the discrepancy is about two
standard deviations. However, if uncertainties on $B_K$ and on CKM parameters
improve substantially, this discrepancy could become significant, as the fifth
point of \fig{fig:epserrors} indicates.

$B_K$ is a quantity which has a long history on and off the lattice. However,
because of space constraints, in \tab{tab:BKparam} I only summarize the
parameters and results of unquenched lattice calculations. The
corresponding consumer report is given in \tab{tab:BKflavia}.

\begin{table}[t]
\begin{center}
\begin{tabular}{lllcccc}
\hline
\hline
Ref. & $N_f$ & action & $a[\fm]$ & $L[\fm]$ & 
\begin{tabular}{c}$M_\pi^{min}[\mev]$\\[-0mm]
typ/val\end{tabular} & 
$\hat B_K$\\
\hline
{JLQCD'08~\cite{Aoki:2008ss}}
& 2 & Overlap
& 0.12 & 1.9 & 290/290 & 0.734(5)(55)\\
{ETM'08~\cite{Dimopoulos:2008hb}}
& 2 & OS/tmQCD & 0.07,0.09 & 2.1,2.7 & 300/300 & 
{0.78(3)}\\
\hspace{-0.2cm}\begin{tabular}{l}
{HPQCD/}\\[-2mm]
{UKQCD'06~\cite{Gamiz:2006sq}}
\end{tabular} & 2+1 & $\mathrm{KS}^\mathrm{HYP}_\mathrm{MILC}$ &
0.125 & 2.5 & 460/360 & 0.85(2)(18)\\
\hspace{-0.2cm}\begin{tabular}{l}
{RBC/UKQCD}\\[-2mm]
{'07-08~\cite{Antonio:2007pb,Allton:2008pn}}
\end{tabular}  
& 2+1 & DWF & 0.11 & 1.8,2.8 & 290/240 & 0.717(14)(35)\\
{Bae et al '08~\cite{Bae:2008tb}}
 & 2+1 & $\mathrm{KS}^\mathrm{HYP}_\mathrm{MILC}$
& $\gsim 0.06$ & 4 & $300/240$ & $\delta \hat B_K\to 3\%$\\
\hline
\hline
\end{tabular}
\end{center}
\caption{\label{tab:BKparam} \em Parameters of the simulations used by various
  collaborations for calculating the renormalization group invariant parameter
  $\hat B_K$, together with their result for that quantity. The values of
  $\hat B_K$ in the table have been obtained at NLO from the RI/MOM or
  $\msbar$-NDR values given in the papers. The description of the columns can
  be inferred from the one given in \protect\tab{tab:FKFpi}. \vspace{-0.3cm}}
\end{table}

\begin{table}[ht]
\begin{center}
\begin{tabular}{l@{\hspace{10mm}}c@{\hspace{10mm}}c@{\hspace{10mm}}c@{\hspace{10mm}}c@{\hspace{10mm}}c@{\hspace{10mm}}c@{\hspace{10mm}}c@{\hspace{14mm}}}
\hline
\hline\\[0.4cm]
Ref. & \begin{rotate}{40}{\footnotesize publication}\end{rotate}
& \begin{rotate}{40}{\footnotesize action, unit.}\end{rotate}
& \begin{rotate}{40}{\footnotesize $N_f$}\end{rotate}
& \begin{rotate}{40}{\footnotesize mass extrap}\end{rotate}
& \begin{rotate}{40}{\footnotesize $a\to 0$}\end{rotate}
& \begin{rotate}{40}{\footnotesize finite volume}\end{rotate}
& \begin{rotate}{40}{\footnotesize renorm}\end{rotate}
\\
\hline
{JLQCD'08~\cite{Aoki:2008ss}} & \good & \good & \soso &  \soso & \bad & \bad & \soso \\
{ETM'08~\cite{Dimopoulos:2008hb}} & \bad & \soso & \soso & \soso  & \soso & \soso & \soso \\
\hspace{-0.2cm}\begin{tabular}{l}
{HPQCD/}\\[-2mm]
{UKQCD'06~\cite{Gamiz:2006sq}}
\end{tabular} & \good & \soso & \good 
& \bad & \bad & \good & \bad \\
\hspace{-0.2cm}\begin{tabular}{l}
{RBC/UKQCD}\\[-2mm]
{'07-08~\cite{Antonio:2007pb,Allton:2008pn}}
\end{tabular} & \good & \good & \good & \soso & \bad & \good & \soso\\
\hline
\hline
\end{tabular}
\end{center}
\caption{\label{tab:BKflavia} \em Starring of the simulations used to
  obtain $B_K$, according to the criteria put forth in
  \protect\sec{sec:stars}. \vspace{-0.3cm}}
\end{table}

At this conference, A. Vladikas presented new, preliminary results for
$B_K$ from ETM~\cite{Dimopoulos:2008hb}. Their calculation makes use of
Osterwalder-Seiler valence quarks on the ETM, $N_f=2$, tmQCD seas to ensure
automatic $O(a)$-improvement as well as multiplicative renormalization of the
$\Delta S=2$, four-quark operator. This implies that their calculation suffers
from $O(a^2)$ unitarity violations, which must be controlled. Their plan is to
use ETM's three lattice spacings, $a\simeq 0.07$, 0.09, 0.010~fm to
extrapolate to the continuum limit. For the moment, though, all results are
obtained from simulations performed at $a \simeq 0.09\,\fm$. They extrapolate
in up and down quark mass using NLO, partially-quenched $SU(2)$ ChPT
\cite{Sharpe:1995qp,Allton:2008pn} and interpolate in valence strange quark
mass linearly. Note that the
extrapolated value of $B_K$ is only $\sim 3\%$ below its value at the lightest
up and down quark mass, suggesting that this extrapolation is well
controlled. The renormalization is performed nonperturbatively in the RI/MOM
scheme~\cite{Donini:1999sf}. The continuum limit and finite-volume corrections
are still missing. Discretization induced unitarity violations will have to be
investigated.

Unquenched, lattice results for $\hat B_K$ are summarized in
\fig{fig:BKsummary}, where my average for this quantity, obtained as explained
in \sec{sec:avg}, is also given. Only the two $N_f=2+1$
results~\cite{Gamiz:2006sq,Antonio:2007pb,Allton:2008pn} are taken into
account in this average. The systematic error is taken
from~\cite{Antonio:2007pb,Allton:2008pn}. However, each calculation was
performed at a single, rather coarse value of the lattice spacing. This
means that these results, and thus the average, suffer from a poorly
controlled discretization errors.

As noted above, the value of $B_K$ has come down quite significantly compared
to JLQCD's standard quenched value of a decade ago~\cite{Aoki:1997nr}. In
particular, $(B_K)_{RBC}^{N_f=2+1}/(B_K)_{JLQCD}^{N_f=0}= 0.83(8)$. This drop
cannot really be ascribed to the inclusion of sea quark effects, since
comparably low results were obtained in the quenched approximation at
comparable lattice spacings~\cite{Aoki:2005ga}. However there, a continuum
extrapolation based on two calculations performed at $a\simeq 0.10\,\fm$ and
0.067~fm increased the result to $(B_K)_{RBC}^{N_f=0}/(B_K)_{JLQCD}^{N_f=0} =
0.90(9)$. Thus, it is very important to clarify this situation by
investigating the continuum limit of $B_K$ in $2+1$ flavor simulations.

The total lattice error on $B_K$ is $\delta B_K^{lat}=5\%$, which is
comparable to the other uncertainties in the standard model prediction for
$\epsilon$. As discussed above, to improve this prediction and possibly reveal
new physics, we must not only reduce the error on $B_K$, but also improve the
determination of CKM parameters, of the correction related to
$\kappa_\epsilon$ and eventually of the short distance QCD coefficients.

\section{Conclusion}

Lattice QCD simulations have made tremendous progress in the last few
years. $2+1$ flavor lattice calculations with pion masses as low as $M_\pi\sim
190\,\mev$ in $(4\,\fm)^3$ volumes, and lattice spacings down to $\sim
0.065\,\fm$ have already been performed~\cite{Durr:2008zz}. Moreover, as
PACS-CS has shown~\cite{Aoki:2008sm}, simulations at physical $M_\pi$ are
around the corner. Thus, it is has now become possible to reach the physical
QCD point ($M_\pi\simeq 135\,\mev$, $a\to 0$, $L\to\infty$) in a controlled
fashion.

Quantities such as $F_K/F_\pi$ and $f_+^{K^0\pi^-}(0)$ are already being
computed with percent or better accuracy and are having an important impact on
SM and BSM tests. Quantities such as $B_K$ are reaching the sub 10\% accuracy
level and have errors which match those from other sources. Calculations of
$\epsilon'/\epsilon$ and the $\Delta I=1/2$ enhancement still have 100\%
uncertainties despite the impressive $N_f=2+1$ RBC/UKQCD
effort~\cite{Li:2008kc}, but perhaps not for long~\cite{Li:2008kc}. Many
quantities are still missing continuum extrapolations.

NLO $SU(3)$ ChPT appear to be having trouble at the physical strange quark
mass, at least in the presence of heavier up and down quarks, whereas $SU(2)$
ChPT performs better. However, these inferences require further investigation,
in particular once continuum limits have been investigated.

Concerning the extrapolations and interpolations required to reach the
physical mass point $(m_{ud},m_s)=(m_{ud}^{ph},m_s^{ph})$, my advice is to
keep an open mind and to pick the approach which gives the lowest combined
statistical and systematic error.

To conclude, the age of precision, nonperturbative QCD calculations is
dawning, and the next few years should bring many exciting results.

\vspace{-0.3cm}
\acknowledgments
\vspace{-0.3cm}
I thank Andrzej Buras, Norman Christ, Diego Guadagnoli, Jack
Laiho, Weonjong Lee, Derek Leinweber, Chris Sachrajda, Enno Scholz, Amarjit
Soni, Cecilia Tarantino, Carsten Urbach and Tassos Vladikas for sharing
results with me and J\'er\^ome Charles, Claude Bernard, Stephan D\"urr, Zoltan
Fodor, Christian Hoelbling, and Steve Sharpe for discussions. This work was
supported in part by the EU network FLAVIAnet MRTN-CT-2006-035482 and by the
CNRS's GDR grant 2921 (Physique subatomique et calculs sur r\'eseau).

\renewcommand{\baselinestretch}{0.}
\bibliographystyle{my-pos}
\bibliography{lat08}

\begin{thebibliography}{}

\bibitem{Aoki:2008sm}
S. Aoki \protect{\it et al.} [PACS-CS], arXiv0807.1661 [hep-lat].

\bibitem{Gasser:1984gg}
J. Gasser and H. Leutwyler, Nucl. Phys. \textbf{B250} (1985) 465.

\bibitem{Gasser:1983yg}
J. Gasser and H. Leutwyler, Ann. Phys. \textbf{158} (1984) 142.

\bibitem{Sharpe:1995qp}
S.R. Sharpe and Y. Zhang, Phys. Rev. \textbf{D53} (1996) 5125-5135,
  arXiv:hep-lat/9510037 .

\bibitem{Allton:2008pn}
C. Allton \protect{\it et al.} [RBC-UKQCD], Phys. Rev. \textbf{D78} (2008)
  114509, arXiv:0804.0473  [hep-lat].

\bibitem{Scholz:2008uv}
E.E. Scholz [RBC], arXiv0809.3251 [hep-lat].

\bibitem{Kuramashi:2008tb}
Y. Kuramashi(2008) , arXiv:0811.2630  [hep-lat].

\bibitem{Hardy:2008gy}
J.C. Hardy and I.S. Towner, arXiv0812.1202 [nucl-ex].

\bibitem{Antonelli:2008jg}
M. Antonelli \protect{\it et al.} [FlaviaNet Working Group on Kaon Decays],
  arXiv0801.1817 [hep-ph].

\bibitem{CKMfitter09}
CKMfitter, update of January 14, 2009 (http://ckmfitter.in2p3.fr/).

\bibitem{Amsler:2008zzb}
C. Amsler \protect{\it et al.} [Particle Data Group], Phys. Lett. \textbf{B667}
  (2008) 1.

\bibitem{Blossier:2008dj}
B. Blossier \protect{\it et al.}, arXiv0810.3145 [hep-lat].

\bibitem{Beane:2006kx}
S.R. Beane \protect{\it et al.}, Phys. Rev. \textbf{D75} (2007) 094501,
  arXiv:hep-lat/0606023 .

\bibitem{Aubin:2004fs}
C. Aubin \protect{\it et al.} [MILC], Phys. Rev. \textbf{D70} (2004) 114501,
  arXiv:hep-lat/0407028 .

\bibitem{Bernard:2007ps}
C. Bernard \protect{\it et al.}, PoS \textbf{LAT2007} (2007) 090,
  arXiv:0710.1118  [hep-lat].

\bibitem{Follana:2007uv}
E. Follana \protect{\it et al.} [HPQCD], Phys. Rev. Lett. \textbf{100} (2008)
  062002, arXiv:0706.1726  [hep-lat].

\bibitem{Marciano:2004uf}
W.J. Marciano, Phys. Rev. Lett. \textbf{93} (2004) 231803, arXiv:hep-ph/0402299
  .

\bibitem{Aubin:2008ie}
C. Aubin \protect{\it et al.}, arXiv0810.4328 [hep-lat].

\bibitem{Durr:2008zz}
S. \protect{D\"urr} \protect{\it et al.}, Science \textbf{322} (2008)
  1224-1227.

\bibitem{Colangelo:2005gd}
G. Colangelo \protect{\it et al.}, Nucl. Phys. \textbf{B721} (2005) 136-174,
  arXiv:hep-lat/0503014 .

\bibitem{Leutwyler:1984je}
H. Leutwyler and M. Roos, Z. Phys. \textbf{C25} (1984) 91.

\bibitem{Ademollo:1964sr}
M. Ademollo and R. Gatto, Phys. Rev. Lett. \textbf{13} (1964) 264-265.

\bibitem{Post:2001si}
P. Post and K. Schilcher, Eur. Phys. J. \textbf{C25} (2002) 427-443,
  arXiv:hep-ph/0112352 .

\bibitem{Bijnens:2003uy}
J. Bijnens and P. Talavera, Nucl. Phys. \textbf{B669} (2003) 341-362,
  arXiv:hep-ph/0303103 .

\bibitem{Jamin:2004re}
M. Jamin \protect{\it et al.}, JHEP \textbf{02} (2004) 047,
  arXiv:hep-ph/0401080 .

\bibitem{Cirigliano:2005xn}
V. Cirigliano \protect{\it et al.}, JHEP \textbf{04} (2005) 006,
  arXiv:hep-ph/0503108 .

\bibitem{Portoles:2007gb}
J. Portoles, arXivhep-ph/0703093.

\bibitem{Tsutsui:2005cj}
N. Tsutsui \protect{\it et al.} [JLQCD], PoS \textbf{LAT2005} (2006) 357,
  arXiv:hep-lat/0510068 .

\bibitem{Dawson:2006qc}
C. Dawson \protect{\it et al.}, Phys. Rev. \textbf{D74} (2006) 114502,
  arXiv:hep-ph/0607162 .

\bibitem{Simula:CKM08}
S. Simula, talk at 5th Workshop on the CKM Unitary Triangle (2008).

\bibitem{Okamoto:2004df}
M. Okamoto [Fermilab Lattice], arXivhep-lat/0412044.

\bibitem{Boyle:2007qe}
P.A. Boyle \protect{\it et al.}, Phys. Rev. Lett. \textbf{100} (2008) 141601,
  arXiv:0710.5136  [hep-lat].

\bibitem{Becirevic:2004ya}
D. Becirevic \protect{\it et al.}, Nucl. Phys. \textbf{B705} (2005) 339-362,
  arXiv:hep-ph/0403217 .

\bibitem{Simula:2007fa}
S. Simula [ETM], PoS \textbf{LAT2007} (2007) 371, arXiv:0710.0097  [hep-lat].

\bibitem{Boyle:2008yd}
P.A. Boyle \protect{\it et al.}, JHEP \textbf{07} (2008) 112, arXiv:0804.3971
  [hep-lat].

\bibitem{Bedaque:2004ax}
P.F. Bedaque and J.W. Chen, Phys. Lett. \textbf{B616} (2005) 208-214,
  arXiv:hep-lat/0412023 .

\bibitem{Flynn:2005in}
J.M. Flynn \protect{\it et al.} [UKQCD], Phys. Lett. \textbf{B632} (2006)
  313-318, arXiv:hep-lat/0506016 .

\bibitem{Guadagnoli:2005be}
D. Guadagnoli \protect{\it et al.}, Phys. Rev. \textbf{D73} (2006) 114504,
  arXiv:hep-lat/0512020 .

\bibitem{Boyle:2007wg}
P.A. Boyle \protect{\it et al.}, JHEP \textbf{05} (2007) 016,
  arXiv:hep-lat/0703005 .

\bibitem{Boyle:2008hd}
P.A. Boyle \protect{\it et al.}, arXiv0812.4265 [hep-lat].

\bibitem{deRafael:1995zv}
E. Rafael, arXivhep-ph/9502254.

\bibitem{Anikeev:2001rk}
K. Anikeev \protect{\it et al.}, arXivhep-ph/0201071.

\bibitem{Buras:2008nn}
A.J. Buras and D. Guadagnoli, Phys. Rev. \textbf{D78} (2008) 033005,
  arXiv:0805.3887  [hep-ph].

\bibitem{Buchalla:1995vs}
G. Buchalla \protect{\it et al.}, Rev. Mod. Phys. \textbf{68} (1996) 1125-1144,
  arXiv:hep-ph/9512380 .

\bibitem{Lunghi:2008aa}
E. Lunghi and A. Soni, Phys. Lett. \textbf{B666} (2008) 162-165,
  arXiv:0803.4340  [hep-ph].

\bibitem{Buras:2009pj}
A.J. Buras and D. Guadagnoli, arXiv0901.2056 [hep-ph].

\bibitem{charles09}
J. Charles, {\protect\it private communication}.

\bibitem{Aoki:2008ss}
S. Aoki \protect{\it et al.} [JLQCD], Phys. Rev. \textbf{D77} (2008) 094503,
  arXiv:0801.4186  [hep-lat].

\bibitem{Dimopoulos:2008hb}
P. Dimopoulos \protect{\it et al.}, arXiv0810.2443 [hep-lat].

\bibitem{Gamiz:2006sq}
E. Gamiz \protect{\it et al.} [HPQCD], Phys. Rev. \textbf{D73} (2006) 114502,
  arXiv:hep-lat/0603023 .

\bibitem{Antonio:2007pb}
D.J. Antonio \protect{\it et al.} [RBC], Phys. Rev. Lett. \textbf{100} (2008)
  032001, arXiv:hep-ph/0702042 .

\bibitem{Bae:2008tb}
T. Bae \protect{\it et al.}, arXiv0809.1220 [hep-lat].

\bibitem{Donini:1999sf}
A. Donini \protect{\it et al.}, Eur. Phys. J. \textbf{C10} (1999) 121-142,
  arXiv:hep-lat/9902030 .

\bibitem{Aoki:1997nr}
S. Aoki \protect{\it et al.} [JLQCD], Phys. Rev. Lett. \textbf{80} (1998)
  5271-5274, arXiv:hep-lat/9710073 .

\bibitem{Aoki:2005ga}
Y. Aoki \protect{\it et al.}, Phys. Rev. \textbf{D73} (2006) 094507,
  arXiv:hep-lat/0508011 .

\bibitem{Li:2008kc}
S. Li and N.H. Christ, arXiv0812.1368 [hep-lat].

\end{thebibliography}

\end{document}